\documentclass[a4paper,aps,prl,twocolumn,showpacs,preprintnumbers,amsmath,amssymb]{revtex4}
 \usepackage{bbm}
 \usepackage{amsmath}
 \usepackage{verbatim}
 \usepackage{graphicx}
 \usepackage{color}
 \usepackage{ulem}
 \bibliography{References}
\begin{document}

 \newcommand{\be}{\begin{eqnarray}}
 \newcommand{\ee}{\end{eqnarray}}
 \newcommand{\bea}{\begin{eqnarray}}
 \newcommand{\eea}{\end{eqnarray}}
 \newcommand{\bma}{\begin{subequations}}
 \newcommand{\ema}{\end{subequations}}
 \newcommand{\bra}[1]{\mbox{$\langle #1 |$}}
 \newcommand{\ket}[1]{\mbox{$| #1 \rangle$}}
 \newcommand{\braket}[2]{\mbox{$\langle #1  | #2 \rangle$}}

 \newenvironment{lemma}[1][]{\begin{trivlist} \it
 \item[\hskip \labelsep {\bfseries #1}]}{\end{trivlist}}

 \newenvironment{proof}[1][]{\begin{trivlist}
 \item[\hskip \labelsep {\bfseries #1}]}{\end{trivlist}}
 \newenvironment{definition}[1][Definition]{\begin{trivlist}
 \item[\hskip \labelsep {\bfseries #1}]}{\end{trivlist}}
 \newenvironment{example}[1][Example]{\begin{trivlist}
 \item[\hskip \labelsep {\bfseries #1}]}{\end{trivlist}}
 \newenvironment{remark}[1][Remark]{\begin{trivlist}
 \item[\hskip \labelsep {\bfseries #1}]}{\end{trivlist}}

 \newcommand{\qed}{\nobreak \ifvmode \relax \else
       \ifdim\lastskip<1.5em \hskip-\lastskip
       \hskip1.5em plus0em minus0.5em \fi \nobreak
       \vrule height0.75em width0.5em depth0.25em\fi}

 \def\1{\ensuremath{\hbox{$\mathrm I$\kern-.6em$\mathrm 1$}}}
 \def\tr{ \mbox{tr}}
 \def\qed{\leavevmode\unskip\penalty9999 \hbox{}\nobreak\hfill
      \quad\hbox{\leavevmode  \hbox to.77778em{%
                \hfil\vrule   \vbox to.675em%
                {\hrule width.6em\vfil\hrule}\vrule\hfil}}
      \par\vskip3pt}

 \def\mark #1 {{\textcolor{red}{#1}}}
 \def\markC #1 {{\textcolor{blue}{#1}}}
 \def\markCh #1 {{\textcolor{green}{#1}}}

 \def\lR{l^2_{\mathbb{R}}}
 \def\RR{\mathbb{R}}
 \def\E{\mathbf e}
 \def\D{\boldsymbol \delta}
 \def\S{{\cal S}}
 \def\T{{\cal T}}
 \def\dd{\delta}
 \def\one{{\bf 1}}
 \def\Flip{{\mathbb{F}}}
 \def\I{{{\mathbb{I}}}}
 \def \eps {\varepsilon}
 \def \tr {\text{tr}}
 \def \ot {\otimes}
 \def \markp #1 {      \begin{itemize}
                          \item \textcolor{red}{#1}
                        \end{itemize}
 }

 \title{Entanglement distillation by dissipation and continuous
 quantum repeaters}

 \author{Karl Gerd H. Vollbrecht$^1$, Christine A. Muschik$^1$, and J. Ignacio Cirac$^1$}

 \affiliation{ $^1$Max-Planck--Institut f\"ur Quantenoptik,
 Hans-Kopfermann-Strasse, D-85748 Garching, Germany
 }

 \begin{abstract}
 Even though entanglement is very vulnerable to interactions with
 the environment, it can be created by purely dissipative
 processes. Yet, the attainable degree of entanglement is
 profoundly limited in the presence of noise sources. We show that
 distillation can also be realized dissipatively, such that a
 highly entanglement steady state is obtained. The schemes put
 forward here display counterintuitive phenomena, such as improved
 performance if noise is added to the system. We also show how
 dissipative distillation can be employed in a continuous quantum
 repeater architecture, in which the resources scale polynomially
 with the distance.
 \end{abstract}

 \pacs{03.67.Ac,03.67.Hk,03.65.Ud}

 \maketitle 

 %
 %
 Entanglement plays a central role in applications of quantum
 information science such as quantum computation, simulation,
 metrology, and communication. However, any quantum technology is
 challenged by dissipation. The interaction of the system with its
 environment is regarded as a major obstacle, and in particular the
 degradation of entangled states due to dissipation is typically
 considered to be a key problem. Contrary to this belief, new
 approaches aim at utilizing dissipation for quantum information
 processes \cite{HAllTheOthers} including quantum state engineering
 \cite{HPoyatosCiracZoller,HKrausZoller1,HFrankWolfCirac}, quantum
 computing \cite{HFrankWolfCirac}, quantum memories
 \cite{HFernando}, the creation of entangled states
 \cite{HScott+EbD_Theory}, and error correction \cite{Hmapo}.

 Entanglement generated by dissipation has been demonstrated
 experimentally \cite{HEbD_Experiment} following a recent
 theoretical proposal \cite{HScott+EbD_Theory}. The main advantage
 of this scheme lies in the fact that entangled states are
 generated in a steady state. Furthermore, as opposed to standard
 methods, the desired state is reached independently of the initial
 one. By coupling two quantum systems to a common environment (e.g.
 the the electromagnetic field \cite{HEbD_Experiment}) a robust
 entangled steady state can be quickly generated and maintained for
 an arbitrary long time without the need for error correction such
 that entanglement is available any time.
 %
 \begin{figure}
 \includegraphics[width=8.5cm]{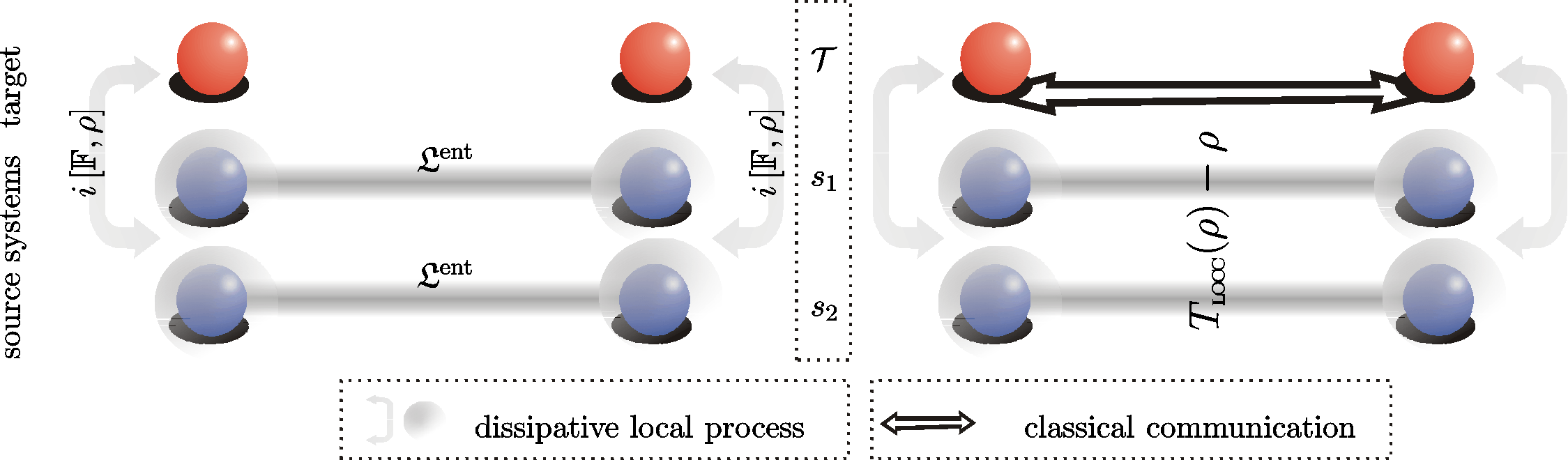}
 \caption{(Color online) Entanglement distillation by dissipation
 a) Distillation setup without communication. b) Distillation setup
 including classical communication.}\label{hugo}
 \end{figure}
 %

 As any other scheme, dissipative protocols are exposed to noise
 sources, which degrade the quality of the produced state and
 render it inapplicable for many important applications in quantum
 information, like quantum communication where noise effects
 increase dramatically with the distance. By means of distillation
 \cite{HDistillation}, entanglement can be improved at the expense
 of using several copies. In combination with teleportation, this
 method allows for the construction of quantum repeaters
 \cite{HBriegel}, which enable the distribution of high-quality
 entanglement for long distance quantum communication with a
 favorable scaling of resources. Unfortunately, existing schemes
 for distillation and teleportation are incompatible with protocols
 generating entanglement in a steady state, since they require the
 decoupling of the system from the environment, such that the
 advantages are lost. Hence, new procedures which are suitable to
 accommodate dissipative methods such that all advantages can be
 retained and used for quantum repeaters are highly desirable.

 We introduce and analyze different dissipatively driven
 distillation protocols, which allow for the production of highly
 entangled steady states independent of the initial one and present
 a novel quantum repeater scheme featuring the same properties.
 More specifically, this protocol continuously produces
 high-quality long-range entanglement. The required resources scale
 only polynomially in the distance. Once the system is operating in
 steady state, the resulting entangled link can be used for
 applications. Remarkably, the time required to drive a new pair
 into a highly entangled steady state is independent of the length
 of the link such that this setup provides a continuous supply of
 long distance entanglement \cite{HBriegel}. Apart from that, the
 proposed distillation protocols exhibit several intriguing
 features. We present, for example, a method which allows for
 distillation in steady state where none of the individual source
 pairs is entangled, and describe another one whose performance can
 be improved by deliberately adding noise to the system.
 %
 %
 %

\begin{figure}
 \includegraphics[width=8.5cm]{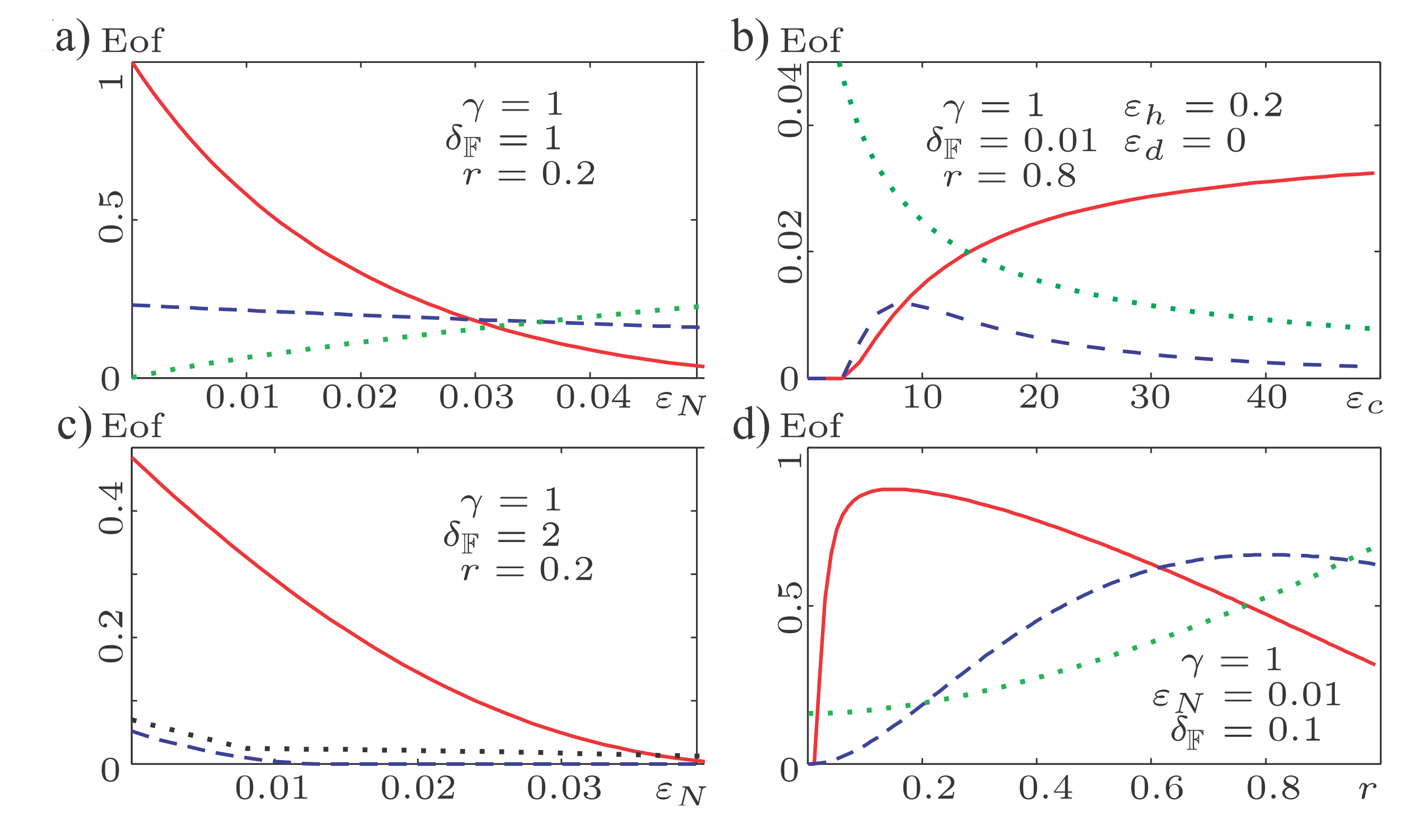}
 \caption{(Color online) Dissipative distillation according to
 scheme I without communication (panel a) and including classical
 communication (panels b-d). The full red lines show the steady
 state entanglement of formation (Eof) of system $\T$. The dashed
 blue lines depict the steady state Eof of the source state s$_1$
 if no distillation is performed (a,c,d) and during the protocol
 (b). For better visibility the blue dashed curve is multiplied by
 a factor 30 in panels b and c. The dotted green lines show the
 entropy of s$_1$ which is a measure of its mixedness. a) EoF
 attainable without communication versus error rate
 $\eps_{\text{N}}\equiv\eps_{\text{h}}=\eps_{\text{c}}=\eps_{\text{d}}$.
 b) EoF versus the noise parameter $\eps_{\text{c}}$. c) EoF versus
 error rate $\eps_{\text{N}}$. The black dotted curve represents
 the entanglement of the total source system measured in log
 negativity. d) EoF versus the parameter $r$ \label{Fig_SchemeI}}.
 \end{figure}

 In the following, we introduce two types of dissipative
 distillation protocols suitable for different situations. We start
 out by explaining scheme I which is physically motivated and
 consider the situation shown in Fig.~\ref{hugo}. Two parties,
 Alice and Bob, share two source qubit pairs s$_1$ and s$_2$, which
 are each dissipatively driven into an entangled steady state and
 used as resource for creating a single highly entangled pair in
 target system $\T$. Assuming Markov dynamics, the time evolution
 of the density matrix $\rho$ can be described by a master equation
 of Lindblad form $\dot{\rho}=\gamma\left(A \rho
 A^\dagger-\frac{1}{2} \left( \rho A^\dagger A + A^\dagger
 A\rho\right)\right)$ with rate $\gamma$ and will be abbreviated by
 the short hand notation $\dot{\rho}=\gamma\mathfrak{L}^{A}(\rho)$
 in the following. The entangling dissipative process acting on the
 source qubits considered here, is the single-particle version of
 the collective dynamics realized in \cite{HEbD_Experiment} and
 corresponds to the master equation
 $\dot{\rho}=\mathfrak{L}^{\text{\tiny{ent}}}(\rho)=\gamma\left(\mathfrak{L}^{A}(\rho)+\mathfrak{L}^{B}(\rho)\right)$
 with $A= \cosh(r) \sigma_{\text{\tiny{Alice}}}^{-}+ \sinh(r)
 \sigma_{\text{\tiny{Bob}}}^{+}$ and $B= \cosh(r)
 \sigma_{\text{\tiny{Bob}}}^{-}+ \sinh(r)
 \sigma_{\text{\tiny{Alice}}}^{+}$, where
 $\sigma^{-}=\ket{0}\bra{1}$ and $\sigma^{+}=\ket{1}\bra{0}$.
 The unique steady state of this evolution is the pure entangled
 state $\ket{\psi}=\left(\ket{00}-\lambda
 \ket{11}\right)/\sqrt{1+\lambda^2}$, where $\lambda=\tanh(r)$. It
 is subject to local cooling, heating and dephasing noise described
 by $ \mathfrak{L}^{\text{noise}}(\rho)= \eps_{\text{c}}
 \left(\mathfrak{L}^{\sigma^{-}_{\!\text{\tiny{Alice}}}}(\rho)\!+\!\mathfrak{L}^{\sigma^{-}_{\!\text{\tiny{Bob}}}}(\rho)\!
 \right)
 \!+\!\eps_{\text{h}}\left(\mathfrak{L}^{\sigma^{+}_{\!\text{\tiny{Alice}}}}(\rho)\!+\!\mathfrak{L}^{\sigma^{+}_{\!\text{\tiny{Bob}}}}(\rho)\!\right)
 \!+\eps_{\text{d}}\left(\mathfrak{L}^{\tiny{\ket{1}\!\bra{1}}_{\!\text{\tiny{Alice}}}}(\rho)\!+\mathfrak{L}^{\tiny{\ket{1}\!\bra{1}}_{\!\text{\tiny{Bob}}}}(\rho)\!\right)
 \label{Eq_NoCom} $.
 %
 %
 %
 %
 We assume that the entangling dynamics acting on s$_1$ and s$_2$
 is noisy, while the target system is protected (this assumption
 will be lifted below). The source qubits are locally coupled to
 $\T$ such that
 $$
 \dot{\rho}\!=\!\mathfrak{L}^{\text{\tiny{ent}}}_{s_1}\!(\!\rho)+\mathfrak{L}^{\text{\tiny{ent}}}_{s_2}\!(\!\rho)+\mathfrak{L}^{\text{\tiny{noise}}}_{s_1}\!(\!\rho)+\mathfrak{L}^{\text{\tiny{noise}}}_{s_2}\!(\!\rho)+\mathfrak{L}_{\text{\tiny{Alice}}}\!(\!\rho)+\mathfrak{L}_{\text{\tiny{Bob}}}\!(\!\rho),
 $$ where $\mathfrak{L}_{\text{\tiny{Alice}}}
 (\mathfrak{L}_{\text{\tiny{Bob}}})$ acts only on Alice's (Bob's)
 side.
 We choose
 $\mathfrak{L}_{\text{\tiny{Alice}}}(\rho)=-\mathfrak{L}_{\text{\tiny{Bob}}}(\rho)=i
 \delta_{\mathbb{F}} \left[\mathbb{F} ,\rho \right]$, corresponding
 to the unitary evolution with respect to the Hamiltonian
 $\mathbb{F} = \sum_{i,j} \ket{j_\text{t} \hat i_{\text{s}}}
 \bra{i_\text{t} \hat j_{\text{s}}}$, where $\ket{\hat 0_{\text{s}}}=\ket{0_{\text{s}_{1}}1_{\text{s}_{2}}}$ and
 $\ket{\hat 1_{\text{s}}}=\ket{1_{\text{s}_{1}}0_{\text{s}_{2}}}$.
 Note that this distillation protocol does not require any
 classical communication or pre-defined correlations. As can be
 seen in Fig.~\ref{Fig_SchemeI}a, the efficiency is mainly
 determined by the mixedness of the source states rather than their
 entanglement. In the absence of errors, the target system reaches
 a maximally entangled state.

 %
 %

\begin{figure}
 \includegraphics[width=7.5cm]{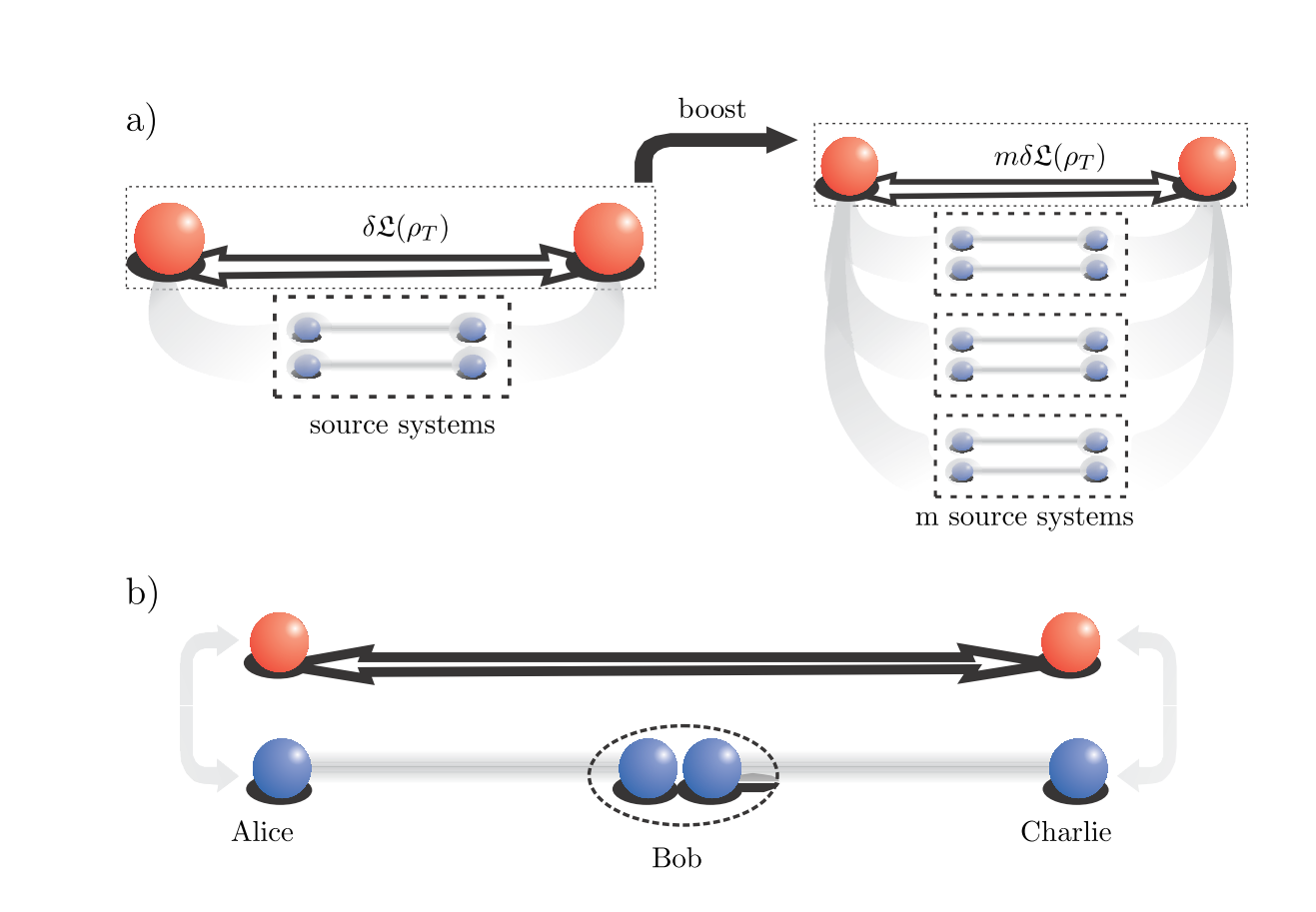}
 \caption{(Color online) Building blocks of a dissipative quantum
 repeater. a) Noise resistent distillation setup. The process
 acting on the target system is boosted using several copies of the
 source system. b) Continuous entanglement swapping procedure.}
 \label{dings}
 \end{figure}

 In order to allow also for noise acting on $\T$, we include now
 classical communication. As shown in the appendix, any Lindblad
 operator of the form
 $ \mathfrak{L}^{T_{\text{\tiny {L\!O\!C\!C}}}}(\rho)\!=\!\left(
 T_{\text{\tiny {L\!O\!C\!C}}}(\rho)-\rho \right),$
 where $T_{\text{\tiny {L\!O\!C\!C}}}$ is an arbitrary LOCC channel
 \cite{HLOCC}, can be realized using local dissipative processes in
 combination with classical communication
 \cite{HRetardationFootnote}.
 In particular, this allows for the stabilization of the
 distillation schemes discussed below against errors acting on the
 target system by running them using $m$ blocks of source pairs,
 which are all coupled to the same target state (see Sec.~3 in
 \cite{supp}) as illustrated in Fig.~\ref{dings}a. If sufficiently
 many source-blocks, $m$, are used, the dynamics is dominated by
 the desired processes. For clarity, we discuss the following
 distillation schemes in the absence of target errors, which
 corresponds exactly to the limit $m\rightarrow\infty$.

 Thus, we consider the master equation
 \bea \dot{\rho}\!=\!\mathfrak{L}^{\text{\tiny{ent}}}_{s_1}(\rho)\!
 +\!\mathfrak{L}^{\text{\tiny{ent}}}_{s_2}(\rho)\!
 +\!\mathfrak{L}^{\text{\tiny{noise}}}_{s_1}(\rho)\!
 +\!\mathfrak{L}^{\text{\tiny{noise}}}_{s_2}(\rho)
 \!+\!\delta_{\mathbb{F}}\left(T_{\mathbb{F}}(\rho)\!-\!\rho\right).
 \nonumber \eea
 The LOOC map $T_{\mathbb{F}}(\rho)$ is defined by the four Kraus
 operators $\Flip_A \ot \Flip_B, P^\bot_A \ot P_B,P_A \ot
 P^\bot_B,P^\bot_A \ot P^\bot_B$, where $P,P^\bot$ are the
 projection onto the one excitation subspace and its orthogonal
 complement. Alice and Bob measure the number of excitations on
 their side. After successful projection onto the subspace with one
 excitation $P_{A}\otimes P_{B}$, a flip operation $\mathbb{F}$ is
 performed, in the unsuccessful case no operation is carried out.

 As shown in Fig.~\ref{Fig_SchemeI}b, the scheme is robust against
 local noise of cooling-type
 ($\mathfrak{L}^{\sigma^{-}}\!\!(\rho)$). This kind of noise can
 even be used to enhance the performance of the distillation
 protocol in the steady state at the cost of a lower convergence
 rate. Thus, counterintiutevly, it can be beneficial to add noise
 to the system in order to increase the distilled entanglement.
 Moreover, the steady state entanglement of the source pairs is
 zero in the absence of cooling noise for the parameters considered
 in Fig.~\ref{Fig_SchemeI}b, if no distillation scheme is
 performed. For increasing $\epsilon_{c}$, the entanglement in
 s$_1$ and s$_2$ increases, reaches an optimal point an decreases
 again. Yet, the entanglement that can be distilled from these
 pairs is monotonously increasing and displays a boost effect.
 Panel c also hints at another counterintuitive effect, namely that
 entanglement can be distilled even though none of the source pairs
 is (individually) entangled in the steady state. This can be
 explained by noticing that the two-copy entanglement can be
 maintained for high noise rates when the single-copy entanglement
 is already vanishing.
 Fig.~\ref{Fig_SchemeI}d shows that the distilled entanglement
 increases considerably for small values of the parameter $r$
 despite the decrease in the entanglement of the source pairs. This
 is due to the fact that the protocol is most efficient for source
 states close to pure states, where it allows one to distill
 quickly highly entangled state.

 In settings where the source states can be highly mixed, another
 distillation scheme (scheme II hereafter) is a method of choice
 and will be explained in the following.
 %
 %
 %
 %
 %
 We analyze a generic model, which can be solved exactly and allows
 one to reduce the discussion to the essential features of
 dissipative entanglement distillation. As in standard distillation
 schemes, we study the general problem in terms of Werner states
 \cite{HWernerStates}, since many situations can be described this
 way and a wide range of processes can be cast in this form by
 twirling \cite{HWernerStates}. Werner states are of the simple
 form $\rho_{\text{W}}(f)=f \Omega +(1-f)(\I-\Omega)/3$, where
 $\Omega$ is a projector onto the maximally entangled state
 $\ket{00}+\ket{11}$, and $\I$ the identity operator.
 %
 %
 We assume a process which drives each source pair into the state
 $\Omega$,
 $\dot{\rho}=\gamma \left(\text{tr}(\rho)\Omega-\rho\right) \equiv
 \gamma E(\rho)$. Local depolarizing noise is added in the form of
 the Lindblad term
 $N(\rho) \equiv\!\left(\rho_{\text{\tiny{Alice}}}\!\otimes\! \1
 \!-\!\rho\right)\!+\!\left(\1\!\otimes\! \rho_{\text{\tiny{Bob}}}
 \!-\!\rho\right),$
 where $\rho_{\text{\tiny{Alice}}}$ ($\rho_{\text{\tiny{Bob}}}$)
 denotes the reduced density matrix of Alice's (Bob's) system and
 $\1$ the normalized identity. This term describes the continuous
 replacement of the initial state by the completely mixed one The
 source system reaches the steady state
 $\rho_s\propto { \gamma \Omega +\eps \1}$
 of the total master equation $ \dot{\rho}= \gamma E(\rho)+
 \frac{\eps}{2} N(\rho)$ at least exponentially fast in $\gamma$
 (see Sec.~4 in \cite{supp}).
 %
 %
 %
 A continuous distillation process based on a standard protocol
 \cite{HWernerDisst} can be constructed considering $n$ source
 pairs which are independently driven into the steady state
 $\rho_s$ and a target system $\T$.
 $\T$ is coupled to the source pairs by a dissipative dynamics of
 the form $\dot{\rho}=\delta_{\text{D}}\left(\text{tr}(\rho)
 T_D(\rho)-\rho\right)$, where the completely positive map
 $T_D(\rho)$ corresponds to a process which acts on the $n$ source
 pairs and distills a single potentially higher entangled copy. The
 output state is written on $\T$, while the $n$ source pairs are
 re-initialized in the state $\1$. The total master equation is
 given by
 \bea \dot{\rho}= \sum_{i=1}^n  \left( \gamma
 E_i(\rho)+\frac{\eps}{2} N_i(\rho) \right)
 +\delta_{D}(T_D(\rho)-\rho),\nonumber \eea
 where $E_i$, $N_i$ denote entangling and noise processes on the
 $i$th source qubit pair.
 The steady state has a fidelity of $ f= \int_0^1 dx  f_D(f_s
 -(f_s-0.25) x^{\frac{\gamma+\eps}{\delta_{\text{\tiny{D}}}}})$,
 where $f_s$ and $f_D(f)=\tr(\Omega T_D(\rho_{\text{W}}(f)^{\ot
 n}))$ are the fidelity of $\rho_s$ and the output of the
 distillation protocol with $n$ input states of fidelity $f$. High
 fidelities require low values of $\delta_{\text{D}}$. However, the
 solution $\rho(t)$ (see \cite{supp}, Sec.~4) shows that fast
 convergence requires high values of this parameter. A low
 convergence speed on the target system is extremely
 disadvantageous if noise is acting on $\T$. Therefore, a boost of
 the process as illustrated in Fig.~\ref{dings}a is required. This
 way, the new convergence rate is given by $m \delta_{\text{D}}$
 while the back action on each source system remains unchanged (see
 \cite{supp}, Sec.~3).
 %
 %
 %

 The distribution of entanglement over large distances is one of
 the big challenges in quantum information science. In quantum
 repeater schemes, entanglement is generated over short distances
 with high accuracy and neighboring links are connected by
 entanglement swapping. This procedure allows one to double the
 length of the links, but comes at the cost of a decrease in
 entanglement for non-maximally entangled states. Therefore a
 distillation scheme has to be applied before proceeding to the
 next stage, which consists again of entanglement swapping and
 subsequent distillation.
 The basic setup for a continuous entanglement swapping procedure
 is sketched in Fig.~\ref{dings}b. It consists of three nodes
 operated by  Alice, Bob and Charlie, where Alice and Bob as well
 as Bob and Charlie share an entangled steady state. By performing
 a teleportation procedure, an entangled link is established
 between Alice and Charlie and written onto the target system,
 while the source systems are re-initialized in the state $\1$.
 This corresponds to LOCC operation $T_{\text{sw}}(\rho)$.
 The whole dynamics is described by the master equation
 $$ \dot{\rho}= \sum_{i=1}^2  \left( \gamma E_i(\rho)+\frac{\eps}{2}
 N_i(\rho) \right) +\delta_{\text{sw}}(T_{\text{sw}}(\rho)-\rho).$$
 The steady state has a target fidelity of $f=\frac{2 \gamma^2}{(2
 \gamma+\delta_{\tiny{sw}})(\gamma+\delta_{\text{\tiny{sw}}})}
 (f_{sw}(f_s)-\frac{1}{4})+ \frac{1}{4} $, where $ f_{sw}(f_s)$ is
 the output fidelity of the entanglement swapping protocol for two
 input states with fidelity $f_s$ (see \cite{supp}, Sec.~5).
 %
 %
 %

\begin{figure}
 \includegraphics[width=8.5cm]{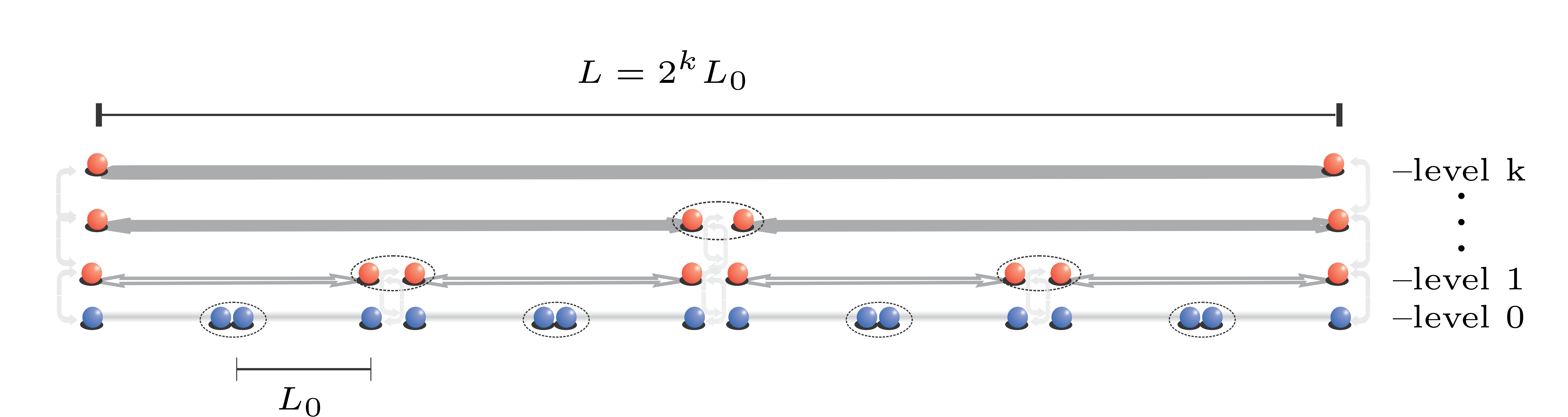}
 \caption{(Color online) Steady state quantum repeater scheme.}
 \label{last}
 \end{figure}

 The basic idea of a nested steady state quantum repeater is
 illustrated in Fig.\ref{last}. At the lowest level, entangled
 steady states are generated over a distance $L_0$. At each new
 level, two neighboring states are connected via a continuous
 entanglement swapping procedure and subsequently written onto a
 target pair separated by twice the distance. The distillation and
 boost processes, that are required in each level to keep the
 fidelity constant are not shown in this picture.
 The resources required for this repeater scheme can be estimated
 as follows. Entanglement swapping processes acting on source pairs
 of length $l$ with fidelity $f_l$ result in entangled target pairs
 of length $2l$, with degraded fidelity $f_{2l}<f_l$. This
 reduction is due to the swapping procedure, noise acting on the
 target system and the back-action from entanglement distillation.
 Stabilization against noise acting on the target systems is
 achieved by coupling each of them to $m$ copies of the source
 system and requires therefore $2m$ source pairs of length $l$. In
 order to obtain a fidelity $f_{2l} \geq f_{l}$, $n$ copies of
 these error stabilized links are used as input for a $n$ to $1$
 distillation process. The distilled state is mapped to another
 target pair of length $2l$, which also needs to be stabilized
 against errors using $m$ copies of the blocks described. Hence, in
 total $2 m^2 n$  pairs of length $l$ are required for a repeater
 stage which doubles the distance over which entanglement is
 distributed. For creating a link of length $L=L_0 2^k$, $(2 m^2
 n)^k$ source pairs are needed, where $k$ is the number of required
 iterations of the repeater protocol. Therefore, the required
 resources scale polynomial with $({L}/{L_0})^{Log_2(2 m^2 n)}$. In
 Sec.~5 in \cite{supp}, we discuss a specific example scaling with
 $(L/L_0)^{16.4}$.
 The convergence time of the total system scales only
 logarithmically with the distance $L$. Once the steady state is
 reached, the entanglement of the last target system can be used
 e.g. for quantum communication or cryptography. The underlying
 source systems are not effected by this process and remain in the
 steady state. Therefore, the target state is restored in constant
 time.
 %
 %

 In conclusion, we have shown how entanglement can be distilled in
 a steady state and distributed over long distances by means of a
 dissipative quantum repeater scheme serving as stepping stone for
 future work aiming at the optimization in view of efficiency and
 experimental implementations.

 \textit{APPENDIX} The continuous exchange of classical
 communication is added in the framework of dissipative quantum
 information processing, by assuming that Alice and Bob have access
 to a system, which is used for communication only and considering
 the master equation
 \begin{eqnarray*}
  \dot \rho\!=\!\Gamma \!\left(\! \sum_{i}\!\!
  \bra{i_{\text{c}_{\text{A}}}}\rho_{\text{Alice}}
 \ket{i_{\text{c}_{\text{A}}}}\ket{0_{\text{c}_{\text{A}}}\!i_{\text{c}_{\text{B}}}}\bra{0_{\text{c}_{\text{A}}}\!i_{\text{c}_{\text{B}}}}\!-\!\rho\!
 \right)\!\equiv \Gamma \mathfrak{C}_{\text{A}\rightarrow
 \text{B}}\!\left(\rho\right)
 \end{eqnarray*}
 States referring to the communication system at Alice's and Bob's
 side are labelled by subscripts c$_\text{A}$ and c$_\text{B}$.
 Alice's communication system is continuously measured in the
 computational basis yielding the quantum state
 $\ket{i_{\text{c}_{\text{A}}}}$ with probability
 $\bra{i_{\text{c}_{\text{A}}}}\rho_{\text{\tiny{Alice}}}\ket{i_{\text{c}_{\text{A}}}}$
 and reset to the state $\ket{0_{\text{c}_{\text{A}}}}$, while the
 communication system on Bob's side is set to the measurement
 outcome. This way, classical information can be sent at a rate
 $\Gamma$, but no entanglement can be created (see \cite{supp},
 Sec.~2).
 As proven in Sec.~2 in \cite{supp}, any operation that can be
 realized by means of local operations and classical communication
 (LOCC) can be constructed in a continuous fashion using
 communication processes $\mathfrak{C}_{\text{A}\rightarrow
 \text{B}}$ and $\mathfrak{C}_{\text{B}\rightarrow \text{A}}$, if
 the rate $\Gamma$ is fast compared to all other relevant processes
 including the retardation due to back and forth communication.
 \\
 We thank Eugene Polzik for helpful discussions and acknowledge
 support from the Elite Network of Bavaria (ENB) project QCCC, the
 DFG-Forschungsgruppe 635 and the EU projects COMPAS and QUEVADIS.

 %
 %
 \newpage
 \setcounter{figure}{0} \setcounter{equation}{0}
 \renewcommand{\thefigure}{S.\arabic{figure}}
 \section{Supplemental Material}

 Here, we explain the results presented in the main text. In Sec.~1
 and Sec.~2, two dissipative distillation schemes are discussed.
 Scheme I is suited for settings where a dissipative processes is
 available which produces entangled steady states, that are close
 to pure states. If only very mixed steady states are available as
 input, scheme II is preferable (which we explain in detail in
 Sec.~4). In Sec.~2, the notion of continuous exchange of classical
 information between two parties is introduced in the master
 equation formalism and it is shown that arbitrary LOCC channels
 can be realized using local dissipation and classical
 communication. In Sec.~3, we explain how the continuous protocols
 used here can be made robust against noise. Finally, in Sec.~5, we
 analyze the dissipative quantum repeater scheme put forward in the
 main text in detail.

 \section{1. Scheme I: Dissipative entanglement distillation for source states close to pure states}
 In this section, we explain two variants of scheme I
 \cite{FootnoteSchemeI}. In Sec.~1.1, we discuss a protocol, which
 allows for dissipative entanglement distillation without
 communication. Sec.~1.2 is concerned with a related protocol,
 which includes classical communication.
 Both protocols produce Bell-diagonal steady states which can
 further distilled using scheme II presented in Sec.~4.

 \subsection{1.1 Dissipative entanglement distillation without communication}
\begin{figure}
 \includegraphics[width=8.5cm]{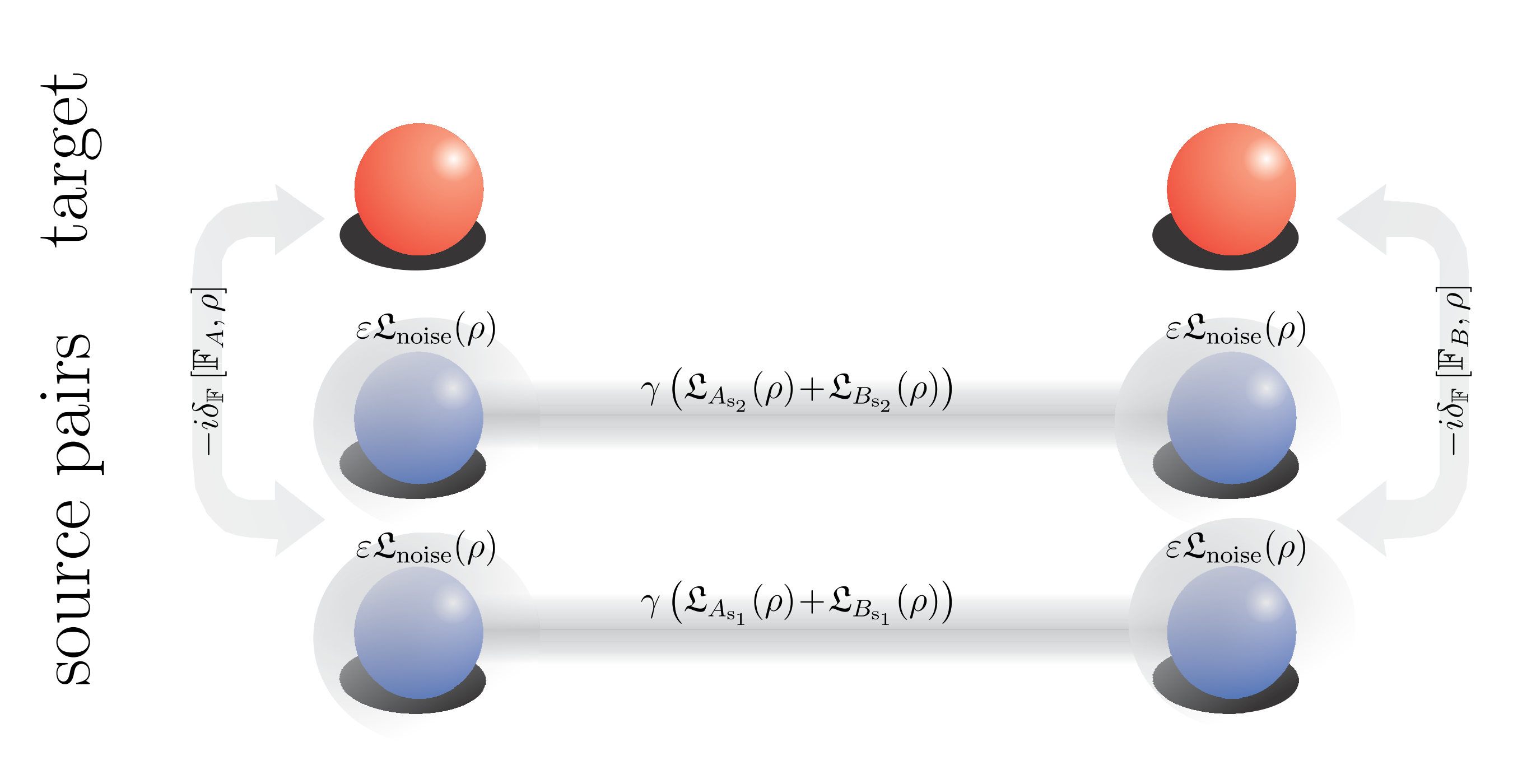}
 \caption{(Color online) Scheme I, dissipative entanglement
 distillation without communication.} \label{ohnebild}
 \end{figure}

 We consider the setup illustrated in Fig.~\ref{ohnebild}. The
 dissipative dynamics driving the two systems $s_1$ and $s_2$ is
 physically motivated and can be implemented by coupling the
 systems located at Alice's and Bob's side to a common bath, for
 example the vacuum modes of the electromagnetic field
 \cite{HScott+EbD_Theory,HEbD_Experiment}. The entanglement which
 can be attained per single copy is limited for a given dissipative
 process. Moreover these systems are subject to noise. Still, it is
 possible to use these two copies as resource for creating a single
 highly entangled pair in target system $\mathcal{T}$.
 In the absence of undesired processes, the dynamics described by
 the master equation
 $\dot{\rho}=\gamma\left(\mathfrak{L}^{A}(\rho)+\mathfrak{L}^{B}(\rho)\right)$
 (see main text) drives systems $s_1$ and $s_2$ into the state
 \bea \ket{\psi}^{\!\ot 2}\!\!\!&\propto&\!\!\!
 \ket{00}_{\!\text{s}\!_1}\!\ket{00}_{\!\text{s}\!_2\!}\!\!-\!\!\lambda\!\left[\ket{00}_{\!\text{s}\!_1}\!\ket{11}_{\!\text{s}\!_2}\!\!+\!\!\ket{11}_{\!\text{s}\!_1}\!\ket{00}_{\!\text{s}\!_2}\right]\!+\!
 \lambda^2
 \ket{11}_{\!\text{s}\!_1}\!\ket{11}_{\!\text{s}\!_2}.\nonumber
 \eea
 Alice and Bob share a maximally entangled state
 $\ket{\Psi_0}\!\!=\!\!\left(\ket{00}_{\!\text{s}\!_1}\!\ket{11}_{\!\text{s}\!_2}\!\!+\!\ket{11}_{\!\text{s}\!_1}\!\ket{00}_{\!\text{s}\!_2}\!\right)/\sqrt{2}$
 in a subspace with one excitation on each side. Scheme I is based
 on the extraction of entanglement from this subspace and its
 subsequent transfer to the target system by means of the flip
 operation $\mathbb{F} = \sum_{i,j} \ket{j_\T \hat {i_{\text{s}}}}
 \bra{i_\T \hat {j_{\text{s}}}}$, where $\ket{\hat 0_{\text{s}}}=\ket{0_{\text{s}_{1}}1_{\text{s}_{2}}}$ and
 $\ket{\hat 1_{\text{s}}}=\ket{1_{\text{s}_{1}}0_{\text{s}_{2}}}$.
 Systems s$_1$ and s$_2$ are permanently driven back to an
 entangled state. In contrast to standard distillation protocols
 for pure states \cite{DisstPure}, the presence of this strong
 process leads to a substantial decrease in the entanglement if the
 flip operations on Alice's and Bob's side are not applied
 simultaneously. Hence, the coordination of their actions, e.g.,
 using fast classical communication, seems to be essential.
 Surprisingly, the desired dynamics can be realized in the absence
 of communication or predefined correlations using local unitary
 evolutions.

 This is possible by exploiting the symmetry of the maximally
 entangled state $\ket{\Psi_0}$. More specifically, $\ket{\Psi_0}$
 is invariant under any unitary operation of the form $U \otimes
 \bar U$, while less entangled pure states are not. $\bar U$
 denotes the complex conjugate of $U$. Such an operation can be
 implemented without communication as the time evolution of a sum
 of local Hamiltonians $H=H_A \ot \I - \I \ot \bar H_B$. Here, we
 use the flip operation such that the corresponding master equation
 is given by
 \begin{eqnarray*}
  \dot\rho &\!=\!&
  \gamma\left(\mathfrak{L}^{A_{\text{s}_1}}(\rho)\!+\!\mathfrak{L}^{B_{\text{s}_1}}(\rho)+\mathfrak{L}^{A_{\text{s}_2}}(\rho)\!+\!\mathfrak{L}^{B_{\text{s}_2}}(\rho)\right)\nonumber\\
  &\!+\!&{i \delta_{\mathbb{F}}} \left[\mathbb{F} \ot \I - \I \ot {\mathbb{F}},\rho
  \right]\nonumber\\
 &\!+\!&\eps_{\text{c}} \left(\mathfrak{L}^{a_{\text{s}_1}}
 (\rho)\!+\!\mathfrak{L}^{b_{\text{s}_1}}
 (\rho)\!+\!\mathfrak{L}^{a_{\text{s}_2}}
 (\rho)\!+\!\mathfrak{L}^{b_{\text{s}_2}} (\rho)\right)
 \nonumber\\
 &\!+\!&\eps_{\text{h}}\left(\mathfrak{L}^{a^{\dag}_{\text{s}_1}}
 (\rho)\!+\!\mathfrak{L}^{b^{\dag}_{\text{s}_1}}
 (\rho)\!+\!\mathfrak{L}^{a^{\dag}_{\text{s}_2}}
 (\rho)\!+\!\mathfrak{L}^{b^{\dag}_{\text{s}_2}} (\rho)\!\right)
 \nonumber\\
 &\!+\!&\eps_{\text{d}}\left(\mathfrak{L}^{a^\dagger_{\text{s}_1}
 a_{\text{s}_1}} (\rho)+\mathfrak{L}^{b^\dagger_{\text{s}_2}
 b_{\text{s}_2}} (\rho)\right),
 \end{eqnarray*}
 where $a=\sigma^{-}_{\!\text{\tiny{Alice}}}$ and
 $b=\sigma^{-}_{\!\text{\tiny{Bob}}}$.
 %
 %
 The first line corresponds to the entangling dissipative process
 (described by nonlocal jump operators $A$ and $B$) acting on the
 two source systems as explained in the main text. The second line
 describes the unitary coupling of the target system to the
 entangled subspace of the two source systems and the last three
 lines represent undesired processes. More specifically, we include
 dephasing at a rate $\epsilon_{\text{d}}$ as well as noise terms,
 which create (annihilate) excitations locally at the heating
 (cooling) rate $\epsilon_{\text{h}}$ ($\epsilon_{\text{c}}$).
 Note that the noise types considered here also include depolarizing noise.
  The
 target system itself is assumed to be protected (below, a variant
 of this scheme is described, which includes classical
 communication and can be made robust against noise acting on the
 target system). \\
 \\A disadvantage of the unitary evolution employed here lies in the
 fact that the source system is subject to a back-action of the
 target state, which depends on the quantum state of $\mathcal{T}$.
 Accordingly, the evolution of the source systems is highly
 dependent on the state of the target pair. It remains an open
 question, whether schemes, similar to the one described in Sec.~3
 can be used to render this protocol robust against errors on the
 target system, or whether this is a special feature of protocols
 including classical communication.
 \subsection{1.2 Distillation using scheme I including classical communication}
 %
 \begin{figure}
 \includegraphics[width=8.5cm]{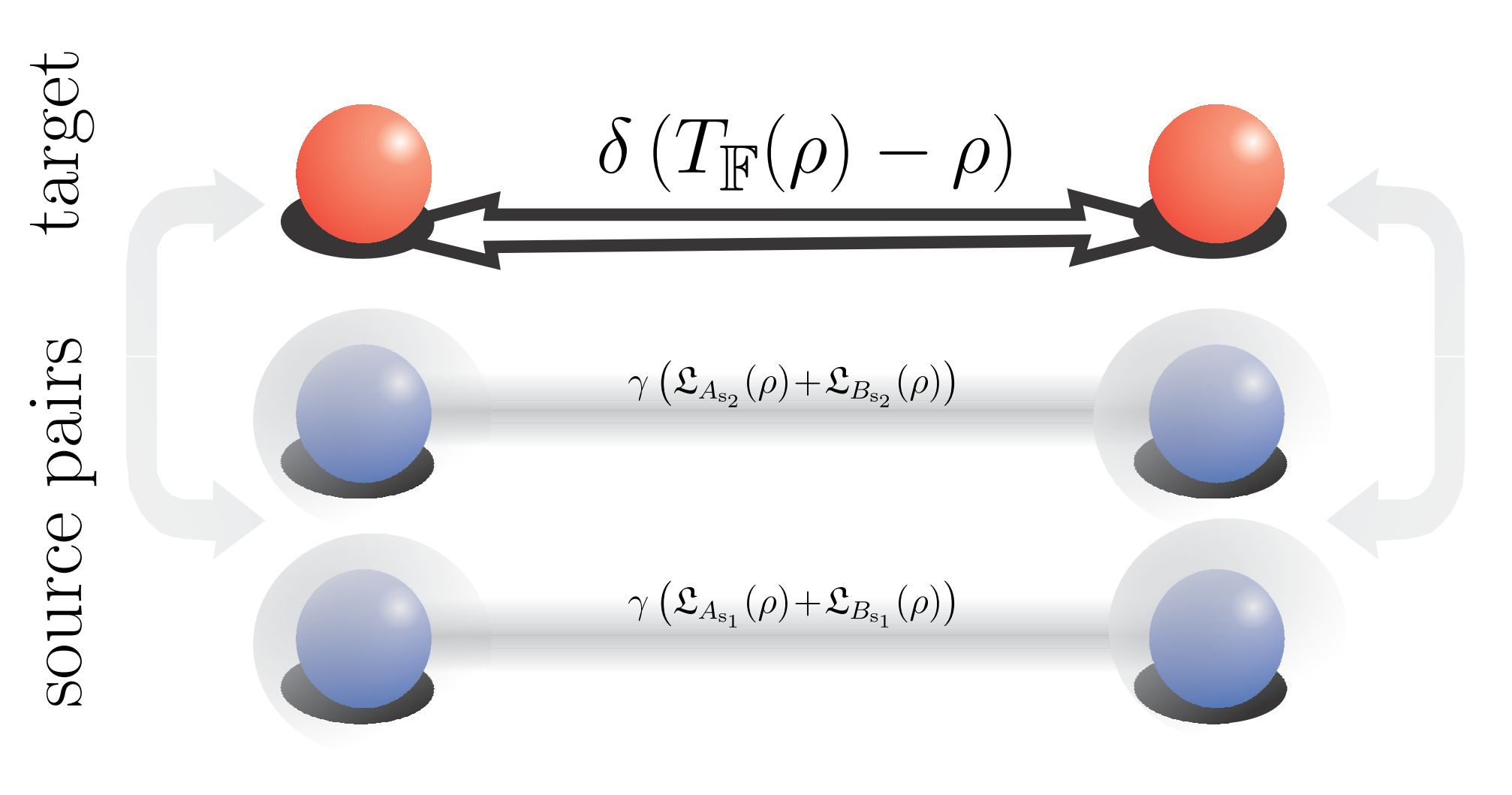}
 \caption{(Color online) Scheme I, dissipative entanglement
 distillation including classical communication
 channels.}\label{Fig:ShemeI:WithCommunication}
 \end{figure}
 %
 We consider the setup illustrated in
 Fig.~\ref{Fig:ShemeI:WithCommunication}. As explained in Sec.~1.1,
 the dissipative entangling process acting on the source systems
 $s_1$ and $s_2$ has the property that Alice and Bob share a
 maximally entangled state
 if the resulting steady state is projected onto the subspace with
 one excitation on each side. Ideally, this quantum state is then
 transferred to the target system $\T$ by means of the flip
 operation defined above. In this subsection, we introduce a
 classical communication channel, which allows Alice and Bob to
 coordinate their actions such that flip operations on both sides
 can be performed in a synchronized fashion if both sides have
 successfully accomplished a projection onto the relevant subspace
 with one excitation.
 As explained in Sec.~2, Lindblad terms of the form $
 \mathfrak{L}^{T_{\text{\tiny {L\!O\!C\!C}}}}(\rho)= \left(
 T_{\text{\tiny {L\!O\!C\!C}}}(\rho)-\rho \right)$, where
 $T_{\text{\tiny {L\!O\!C\!C}}}$ is an arbitrary LOCC channel
 \cite{HLOCC}, can be realized by means of local dissipative
 processes and classical communication. \\
 As explained in Sec.~3, this protocol is resistent against target
 errors if it is coupled to sufficiently many blocks source pairs.
 For simplicity, we explain here the basic protocol in the absence
 of target errors, which corresponds to the limit of using
 infinitely many source blocks (entanglement distillation for a
 finite number of source blocks and finite error rates is analyzed
 in Sec.~3 and Sec.~5).
 Classical communication allows for the implementation of the
 scheme outlined above. The LOCC distillation operation
 corresponding to this process, $T_\Flip$, is given by
 \begin{eqnarray*}
 T_\Flip(\rho)&=&\Flip_\text{{A}}\ot\Flip_\text{{B}} \rho \Flip_\text{{A}}\ot\Flip_\text{{B}}\\
 \nonumber &&+P_A\ot \ P^\bot_B \rho P_A\ot P^\bot_B  \\ \nonumber
 &&+P^\bot_A\ot {P}_B \rho P^\bot_A\ot {P}_B  \\ \nonumber
 &&+P^\bot_A\ot P^\bot_B \rho P^\bot_A\ot P^\bot_B,  \nonumber
 \end{eqnarray*}
 where $P=\ket{0_{s_1} 1_{s_2}}\bra{0_{s_1} 1_{s_2}}+\ket{1_{s_1} 0_{s_2}}\bra{1_{s_1} 0_{s_2}}$ is the projector onto the subspace
 with one excitation, and $P^\bot=\1-P$ the projector onto the
 subspace with zero or two excitations. Note that only the first
 term has an effect on the target system. The flip operation leads
 to a back-action on the source system, which depends on the state
 of $\T$. In order to simplify the discussion in Sec.~3, we
 introduce here a slightly modified version of this protocol,
 $T'_\Flip(\rho)$, which does not exhibit a state-dependent
 back-action. This can be avoided by applying a twirl
 \cite{HWernerStates} on the target system prior to the flip
 operation
 \begin{eqnarray*}
 T'_\Flip(\rho)&=&\sum_{ij=0}^3 \frac{1}{16}\Flip_\text{{A}}\ot\Flip_\text{{B}} U_{ij} \rho U^\dagger_{ij} \Flip_\text{{A}}\ot\Flip_\text{{B}}\\
 \nonumber &&+P_A\ot \ P^\bot_B \rho P_A\ot P^\bot_B  \\ \nonumber
 &&+P^\bot_A\ot {P}_B \rho P^\bot_A\ot {P}_B  \\ \nonumber
 &&+P^\bot_A\ot P^\bot_B \rho P^\bot_A\ot P^\bot_B. \nonumber
 \end{eqnarray*}
 $U_{ij}=\sigma^A_i \ot \sigma_j^B$ is a unitary operation acting
 on the target system only, where $\sigma_i$ denote the four Pauli
 matrices and $\sigma_0$ is the identity. Due to the twirl, this
 protocol features an enhanced back-action on the source, which is
 independent from the target. It turns out that the performance of this
 protocol is qualitatively the same as shown in Fig.~2 in the main
 text. The total master equation is then given by
 \begin{eqnarray*}
  \dot\rho &\!=\!&
  \gamma\left(\mathfrak{L}^{A_{\text{s}_1}}(\rho)\!+\!\mathfrak{L}^{B_{\text{s}_1}}(\rho)+\mathfrak{L}^{A_{\text{s}_2}}(\rho)\!+\!\mathfrak{L}^{B_{\text{s}_2}}(\rho)\right)\nonumber\\
  &\!+\!&{ \delta_{\mathbb{F}}} \left( T_\Flip'(\rho)-\rho \right)\nonumber\\
 &\!+\!&\eps_{\text{c}} \left(\mathfrak{L}^{a_{\text{s}_1}}
 (\rho)\!+\!\mathfrak{L}^{b_{\text{s}_1}}
 (\rho)\!+\!\mathfrak{L}^{a_{\text{s}_2}}
 (\rho)\!+\!\mathfrak{L}^{b_{\text{s}_2}} (\rho)\right)
 \nonumber\\
 &\!+\!&\eps_{\text{h}}\left(\mathfrak{L}^{a^{\dag}_{\text{s}_1}}
 (\rho)\!+\!\mathfrak{L}^{b^{\dag}_{\text{s}_1}}
 (\rho)\!+\!\mathfrak{L}^{a^{\dag}_{\text{s}_2}}
 (\rho)\!+\!\mathfrak{L}^{b^{\dag}_{\text{s}_2}} (\rho)\!\right)
 \nonumber\\
 &\!+\!&\eps_{\text{d}}\left(\mathfrak{L}^{a^\dagger_{\text{s}_1}
 a_{\text{s}_1}} (\rho)+ \mathfrak{L}^{b^\dagger_{\text{s}_2}
 b_{\text{s}_2}} (\rho)\right), \label{Eq_NoCom}
 \end{eqnarray*}
 where $a=\sigma^{-}_{\!\text{\tiny{Alice}}}$ and
 $b=\sigma^{-}_{\!\text{\tiny{Bob}}}$.

 \section{2. Classical dissipative channels and dissipative LOCC }
 Classical channels are easier to realize experimentally than their
 quantum counterparts and can for example be implemented using
 optical fibers. Since classical channels are insufficient for the
 generation of quantum correlations, long-range links can be
 established over large distances using the toolkit of classical
 error-correction. The class of LOCC operations, i.e. quantum
 operations that can be performed using local operations and
 classical communication, is of essential importance in quantum
 information theory, especially in the context of  entanglement
 distillation protocols.

 In this section, we introduce the notion of classical channels in
 the framework of dissipative quantum information processing. This
 allows us to formulate generalized LOCC operations in a continuous
 dissipative setting, which includes a wide range of continuous
 distillation protocols.
 \subsection{2.1 Classical dissipative channels}
 We start out by introducing a dissipative classical communication
 channel.
 %
 \begin{figure}
 \includegraphics[width=8.5cm]{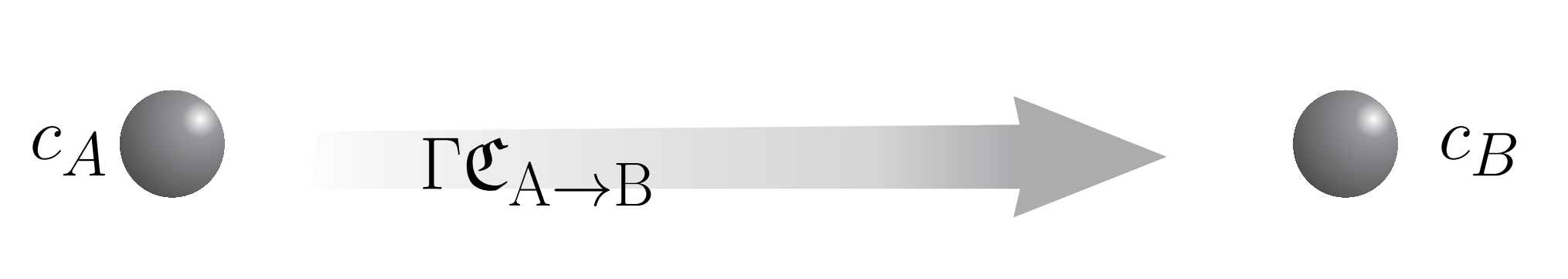}
 \caption{(Color online) Realization of a classical dissipative
 channel.} \label{fig:classq}
 \end{figure}
 %
 Both parties, Alice and Bob, each have access to a $d$-dimensional
 system which is used exclusively for classical communication (see.
 Fig.~\ref{fig:classq}). The master equation
 \begin{eqnarray*}
  \dot \rho\!=\!\Gamma \!\left(\! \sum_i\!
  \bra{i_{\text{c}_{\text{A}}}}\rho_{\text{Alice}}
 \ket{i_{\text{c}_{\text{A}}}}\ket{0_{\text{c}_{\text{A}}}i_{\text{c}_{\text{B}}}}\bra{0_{\text{c}_{\text{A}}}i_{\text{c}_{\text{B}}}}\!-\!\rho\!
 \right)\!\!\equiv\! \Gamma \mathfrak{C}_{\text{A}\!\rightarrow\!
 \text{B}}\!\left(\rho\right)\!.\nonumber
 \end{eqnarray*}
 describes a one-way classical communication channel. States
 referring to the communication system at Alice's and Bob's side
 are labelled by subscripts c$_\text{A}$ and c$_\text{B}$
 respectively. Alice's communication system is continuously
 measured in the computational basis yielding the quantum state
 $\ket{i_{\text{c}_{\text{A}}}}$ with probability
 $\bra{i_{\text{c}_{\text{A}}}}\rho_{\text{\tiny{Alice}}}\ket{i_{\text{c}_{\text{A}}}}$
 and reset to the state $\ket{0_{\text{c}_{\text{A}}}}$, while the
 communication system on Bob's side is set to the measurement
 outcome. This process can be written in the form
 \begin{eqnarray*}
 \dot{\rho}= \Gamma \mathfrak{C}_{\text{A}\rightarrow
 \text{B}}\!\left(\rho\right) \equiv \Gamma(T(\rho)-\rho)
 ,\nonumber
 \end{eqnarray*}
 where the completely positive map $T(\rho)$ is an entanglement
 breaking operation \cite{breaking}, which maps any state to a
 separable one. The solution of this master equation $\rho(t)$ is
 given by
 \begin{eqnarray*}
 \rho(t)=\rho(0) e^{- \Gamma t}+\underbrace{\int_0^t d\tau
 T(\rho(\tau)) e^{\Gamma(\tau-t)}}_{\text{separable}}.
 \end{eqnarray*}
 The second term is separable, since $T(\rho)$ is entanglement
 breaking. Accordingly, the classical channel introduced above does
 not produce entanglement. Moreover any entanglement present in the
 state $\rho(0)$ is exponentially suppressed.

 \subsection{2.2 Generation of Lindblad operators of the form $T(\rho)-\rho$}
\begin{figure}
 \includegraphics[width=8.5cm]{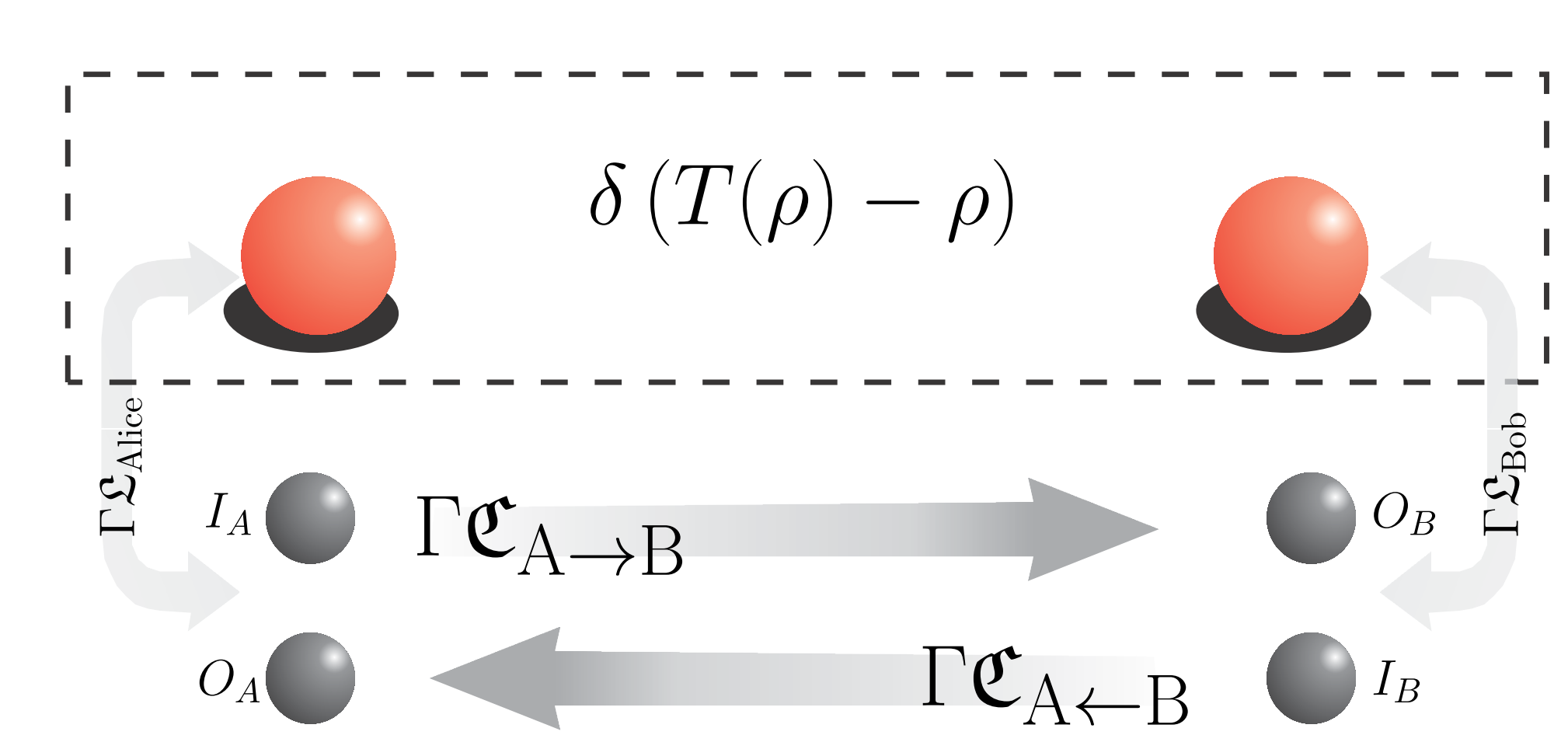}
 \caption{(Color online) Using local dissipative processes and fast
 classical communication, arbitrary LOCC channels can be
 implemented in a continuous fashion.} \label{fig:class}
 \end{figure}

 In the following we prove that any dissipative time evolution
 which satisfies a master equation of the form $\dot
 \rho=\gamma(T(\rho)-\rho)$ can be designed by means of local
 dissipative processes in combination with the classical
 communication channels introduced above in the limit of high rates
 $\Gamma$. The basic setup is sketched in Fig.~\ref{fig:class}.
 Alice and Bob hold a bipartite system, which we refer to as the
 main system. In addition both parties have access to several
 classical communication channels and can apply dissipative
 dynamics acting on the classical channels and their part of the
 main system. This setting allows for a wider class of dissipative
 evolutions on the main system which includes dissipative LOCC
 processes. In particular we state the following:
 \begin{lemma}
 Let $T(\rho)$ be any LOCC map. Let $\mathfrak{L}(\rho)$  be any
 bounded Lindblad operator, i.e., $ \max_\rho \| \mathfrak{L}(\rho)
 \|=1$, acting on the main system at a rate $\gamma$. Let Alice and
 Bob have access to classical communication channels as described
 above. If both parties can apply any dissipative process of
 Lindblad form on their side, an effective dissipative time
 evolution on the main system satisfying the master equation
 \begin{eqnarray}\label{Eq:TheoremME}
 \dot \rho=\gamma \mathfrak{L}(\rho)+\delta \left(T'(\rho)-\rho
 \right)
 \end{eqnarray}
 after an initial waiting time of the order $\frac{1}{\delta}$ can
 be realized. The completely positive operator
 $T'(\rho)\!=\!T(\rho)\!+\!\mathcal{O}(\sqrt{\alpha})$ is an
 imperfect realization of $T(\rho)$ up to an error
 $\mathcal{O}(\sqrt{\alpha})$, which vanishes for small
 $\alpha\!=\!\frac{\gamma'}{\Gamma}$, where
 $\gamma'\!=\!\gamma\!+\!\delta$. $\mathcal{O}(f(\alpha))$ denotes
 any hermitian (time and state dependent) operator with a trace
 norm scaling with $f(\alpha)$ in the limit $\alpha\!\rightarrow\! 0$.
 \end{lemma}
 Since $\max_\rho \| \mathfrak{L}(\rho) \|\!=\!1$, the strength of
 the process is completely encoded in $\gamma$.
 $\mathfrak{L}(\rho)$ can include a dissipative LOOC map itself, as
 discussed at the end of this section. The error
 $\mathcal{O}(\sqrt{\alpha})$ of the LOCC map is small for
 $\alpha\!\ll\!1$.
 A time evolution satisfying Eq.~(\ref{Eq:TheoremME}) can be either
 obtained by starting from certain initial conditions, or after an
 initial waiting time on the order of $\frac{1}{\delta}$, during
 which no external control is required.
 If $\Gamma\!\gg\!\gamma\!+\!\delta$ ($\alpha\!\ll\! 1$), the
 system evolves approximately according to $\dot \rho\!=\!\gamma
 \mathfrak{L}(\rho)\!+\!\delta \left(T(\rho)\!-\!\rho \right)$.\\
 %
 %
 %
 %
 Note, that LOOC operations are extremely hard to parameterize. It
 is known, that they can be written as a separable superoperator
 $T(\rho)=\sum_i A_i \ot B_i \rho A_i^\dagger \ot B_i^\dagger$, but
 not every separable superoperator is a LOCC map. Practically, a
 general LOCC map can only be characterized by fixing the number of
 communication rounds between Alice and Bob and to specifying the
 exact operations that Alice and Bob perform in each round. The
 most general operation Alice and Bob can apply is a positive
 operator valued measurement (POVM). This covers any completely
 positive map as well as measurements, unitary evolutions, etc. A
 POVM is specified by a number of Kraus operators $A_i$,
 corresponding to the possible measurement outcomes $i$, where the
 normalization condition $\sum_i A^\dagger_i A_i=\I$ guaranties
 that the probabilities for the different possible outcomes add up
 to $1$.

 We consider the following situation. Alice performs a first POVM
 $A_i$ and sends her result $i$ to Bob. Bob chooses a POVM
 $B_j^{i}$ depending on Alice's result $i$. Subsequently, he sends
 the result $j$ to Alice, who chooses her next POVM $A_k^{ij}$
 which may depend on all previous measurement result. This
 procedure can be repeated many times. This corresponds to the
 application of the operation
 \begin{eqnarray*}
 T(\rho) \!=\!\!\!\!\!\!\!\!\sum_{i_0,j_0, i_1,j_1,\dots,
 i_r,j_r}\!\!\!\! \!\!\!\!X_{i_0,j_0, i_1,j_1,\dots, i_r,j_r} \rho
 X_{i_0,j_0, i_1,j_1,\dots, i_r,j_r}^\dagger,
 \end{eqnarray*}
 where each Kraus operator is of the form
 \bea \nonumber X_{i_0,j_0, i_1,j_1,\dots, i_r,j_r}=B^{i_0,j_0,i_1
 \dots i_n}_{j_n} \dots B^{i_0,j_0,i_1}_{j_1} A^{i_0,j_0}_{i_1}
 B_{j_0}^{i_0} A_{i_0} \eea
 and represent one possible set of measurements outcomes ${i_0,j_0,
 i_1,j_1,\dots, i_r,j_r}$ for all $r$ POVM measurements. Due to
 this lack of a concise notation for a general LOCC map, a complete
 proof of this statement would be lost in notation and it would be
 hard for the reader to understand the main idea of the proof. We
 restrict ourselves therefore to LOCC maps with one communication
 round, i.e., Alice sends one message to Bob and Bob can send an
 answer back to Alice once.
 A generalization of the following proof to a LOCC map with a
 finite number of communication rounds $m$ is straight forward and
 will be discussed below.\\
 \\Let $T(\rho)$ denote a LOCC map with one round of
 communication. This map can be realized in the following way:
 \begin{itemize}
   \item Alice applies a POVM measurement with Kraus operators $A_i$, obtains the measurement result $i$ and sends it to Bob.
   \item Bob performs a POVM measurement $B_j^i$, which can depend on $i$, and sends the result
   $j$ to Alice. Since we assume that Alice is memoryless, Bob also sends the measurement outcome of Alice's measurement $i$.
   \item In the last step, Alice can apply any completely positive map $T_{ij}$ on her side. This map can depend on both, $i$ and $j$.
 \end{itemize}
 Note, that $T_{ij}(\rho)=\sum_{k} C^{ij}_k \rho
 (C_k^{ij})^\dagger$ is also a POVM map with Kraus operators
 $C^{ij}_k$ , where the measurement results $k$ are not used. Let
 Alice and Bob have $n$ different measurement results for each
 POVM, where $n$ can be upper bounded by the square of the
 dimension of the system. For typical distillation protocols on
 qubits, $n=2$. Note that all indices for Kraus operators run from
 $1$ to $n$, and do not start with $0$. This choice allows for a
 shorter notation later on (the index $0$ is reserved for
 indicating that the classical channel is operable). The basic
 setup is illustrated in Fig.~\ref{fig:class}. Alice and Bob have
 access to classical one-way communication channels labelled $C1$
 and $C2$. $C1$ and $C2$ can be used to send information from Alice
 to Bob and vice versa respectively. Apart from these classical
 channels, Alice and Bob hold a system subject to a dissipative
 evolution described by the Lindblad operator $\mathfrak{L}(\rho)$.
 In the following, this system is referred to as the main system.
 The first classical channel needs to store all possible
 measurement outcomes obtained by Alice, whereas the second one
 needs to store the measurement results obtained by both, Alice and
 Bob. We assume therefore that $C1$ and $C2$ are $n+1$ and
 $n'+1=n^2+1$ dimensional systems respectively. Note, that the
 state $\ket{0}$ will be used to indicate that the channel input or
 output is "empty", while the states $\ket{1},\dots,\ket{n}$
 represent $n$ possible measurement results of Alice and the states
 $\ket{1},\dots,\ket{n'}$ encode the $n^2$ different measurement
 results obtained by Alice and Bob. The corresponding master
 equation is given by
 \begin{eqnarray*}
 \dot \rho=\gamma \mathfrak{L}(\rho)+\Gamma
 \mathfrak{C}_{{\text{A}\rightarrow \text{B}}}(\rho)+\Gamma
 \mathfrak{C}_{{\text{A}\leftarrow \text{B}}}(\rho),
 \end{eqnarray*}
 where $\gamma \mathfrak{L}(\rho)$ is a process acting on the main
 system only. We assume, that the time scales for classical
 communication $\Gamma^{-1}$  are sufficiently long such that
 retardation effects can be ignored. The four systems used for
 classical communication are denoted by $I_a,I_b,O_a,O_b$ as shown
 in Fig.~\ref{fig:class}. $I$ and $O$ stand for "Input" and
 "Output".

 As a next step, local Lindblad operators are added, which
 correspond to the application of a LOCC map depending on the
 registers of the classical channels. The following three terms are
 added, one for each step of the protocol outlined above. The first
 term is given by,
 \begin{eqnarray*} \delta \sum_{i=1,k=0} \mathfrak{L}^{A_i\ot
 \ket{i}\bra{k}_{I_a} }(\rho),
 \end{eqnarray*}
 where $A_i$ acts on Alice's part of the main system and
 $\ket{i}\bra{0}_{I_a}$ on Alice's side of the first classical
 system, i.e., the input system of the first classical channel.
 Note, that we use here the short hand notation
 $\mathfrak{L}^A(\rho)=\gamma\left(A \rho A^\dagger-\frac{1}{2}
 \left( \rho A^\dagger A + A^\dagger A\rho\right)\right)$ that was
 already introduced in the main text. This corresponds to the first
 step of the realization of the LOCC map. Alice performs a POVM and
 writes the measurement result onto the input system of the
 classical channel. As second term, we add the Lindblad operator
 \begin{eqnarray*} \Gamma \sum_{ji=1,xy=0} \mathfrak{L}^{B^i_j\ot
 \ket{0}\bra{i}_{O_b} \ot \ket{j,i}\bra{x,y}_{I_b} }(\rho)
 \end{eqnarray*}
 where $B^i_j$ acts on Bob's part of the main system,
 $\ket{0}\bra{i}_{O_b}$ on the output of the first classical
 channel and $\ket{j,i}\bra{x,y}_{I_b}$ on the input of the second
 channel. Note, that the second channel can store both values $i$
 and $j$ at the same time. $\ket{j,i}$ stands for any encoding of
 $i,j$ in the $n^2+1$ dimensional state space, where the label zero
 is reserved for indicating the status of the channel. The
 summation over $x,y$ starts from zero, i.e., includes the reserved
 zeros term as well as the $n^2$ possible measurement results. Bob
 only carries out a POVM measurement, if he receives the message
 $i$ via $C1$. Afterwards he writes $i,j$ onto the classical
 channel. Note that the sum over $xy$ implies that Bob overwrites
 any previous state of the classical communication system. The last
 term to be added is given by
 \begin{eqnarray*}
 \Gamma \sum_{jik=1} \mathfrak{L}^{C^{ij}_k\ot
 \ket{0}\bra{i,j}_{O_a} }(\rho)
 \end{eqnarray*}
 where $C^{ij}_k$ acts on Alice's quantum system with
 $T_{ij}(\rho)=\sum_k C^{ij}_k\rho {C^{ij}_k}^\dagger $ and
 $\ket{0}\bra{i,j}_{O_a}$ act on the output of the second classical
 channel. Alice receives the message $ij$ and reacts by applying
 $T_{ij}$ to complete the LOCC map. The sum starts from $ij=1$,
 i.e., Alice acts only if a message has arrived. She does not act
 if the register is empty ($\ket{0}$). Hence, the total master
 equation is given by
 \bea \label{eq:total}\dot
 \rho&=&\gamma \mathfrak{L}(\rho)+\Gamma
 \mathfrak{C}_{{\text{A}\rightarrow \text{B}}}+\Gamma
 \mathfrak{C}_{{\text{A}\leftarrow \text{B}}}
 \\\nonumber &&\delta \sum_{i=1,k=0} \mathfrak{L}^{A_i\ot
 \ket{i}\bra{k}_{I_a}  } (\rho)
 \\ \nonumber &&
 +\Gamma \sum_{ji=1,xy=0} \mathfrak{L}^{B^i_j\ot
 \ket{0}\bra{i}_{O_b} \ot \ket{j,i}\bra{xy}_{I_a} } (\rho)
 \\ \nonumber &&+\Gamma \sum_{jik=1}
 \mathfrak{L}^{C^{ij}_k\ot \ket{0}\bra{i,j}_{O_a}  }(\rho) \eea
 The basic idea can be described as follows. The term in the second
 line starts the process of realizing $T(\rho)$ at a rate $\delta$
 (Alice performs the first step). The following steps are performed
 with a high rate $\Gamma$, such that the state of the quantum
 system stays approximately constant during the time needed to
 complete the whole operation. So, practically, the whole LOCC map
 $T(\rho)$ is applied at once at a rate $\delta$.\\
 In the following, this will be proven rigorously by considering
 the effective evolution of the main systems after tracing out the
 classical channels. The reduced state of the main system can be
 written as
 \begin{eqnarray*}
 \rho_M=\sum_{ijkl} \rho_{ijkl}
 \end{eqnarray*}
 with $\rho_{ijkl}=\bra{i_{I_a}j_{O_b}k_{I_b}l_{O_a}}
 \rho\ket{i_{I_a}j_{O_b}k_{I_b}l_{O_a}}$, where $i$, $j$ ($k$, $l$)
 denote the computational bases for $C1$ ($C2$). The indices are
 arranged such that their order corresponds to the order in the
 communication cycle. $i$ refers to the input of Alice's side, $j$
 to the output on Bob's side, $k$ to the input on Bob's side and
 $l$ to the output on Alice's side. A system of differential
 equations for all $\rho_{ijkl}$ can be derived using
 $\dot{\rho}_{ijkl}=\bra{i_{I_a}j_{O_b}k_{I_b}l_{O_a}}
 \dot{\rho}\ket{i_{I_a}j_{O_b}k_{I_b}l_{O_a}}$ and
 Eq.~(\ref{eq:total}).
 The desired terms
 $\rho_{0000},\rho_{i000},\rho_{0i00},\rho_{00(ij)0},\rho_{000(ij)}$
 evolve according to
 \begin{eqnarray}
 \dot{\rho}_{0000}\!\!\!&=&\!\!\!\gamma \mathfrak{L}(\rho_{0000})
 -\delta {\rho}_{0000} +\Gamma \sum_{xy=1}
 T_{ij}({\rho}_{000(xy)}),
 \\\nonumber \label{eq:x000}
 \dot{\rho}_{i000}\!\!\!&=&\!\!\! \gamma
 \mathfrak{L}(\rho_{i000})-(\Gamma+\delta) {\rho}_{i000}+\delta
 \sum_{k=0}  A_i {\rho}_{k000} A_i^\dagger\\ &+&\Gamma \sum_{xy=1}
 T_{xy}({\rho}_{i00(xy)}),
 \\ \label{eq:0x00}
 \dot{\rho}_{0i00}&=&\gamma
 \mathfrak{L}(\rho_{0i00})-(\Gamma+\delta) {\rho}_{0i00}+\Gamma
 \sum_{k=0}{\rho}_{ik00}\nonumber \\
 \!\!\!&+&\!\!\!\Gamma \sum_{xy=1} T_{xy}({\rho}_{0i0(xy)}),
 \\  \label{eq:00x0}
 \dot{\rho}_{00(ij)0}\!\!&=&\!\!\gamma
 \mathfrak{L}(\rho_{00(ij)0})-(\Gamma+\delta) {\rho}_{00(ij)0}
 \\\nonumber \!\!\!&+&\!\!\!\Gamma \!\!\! \sum_{xy=0} \!\!\!B_j^i {\rho}_{0i(xy)0}
 B_j^i{^\dagger}
  \!\!+\!\!\Gamma \sum_{xy=0} T_{xy}({\rho}_{00(ij)(xy)}),
 \\ \label{eq:000x}
 \dot{\rho}_{000(ij)}\!\!\!&=&\!\!\!\gamma
 \mathfrak{L}(\rho_{000(ij)}\!)\!\!-\!\!(\Gamma\!\!+\!\!\delta)
 {\rho}_{000(ij)}\!\!+\!\!\Gamma \!\!\!\sum_{xy=0}\!\!\!
 {\rho}_{00(ij)(\!xy)}\!.
 \end{eqnarray}
 All other terms correspond to small errors. In the first step of
 the proof it is shown that after an initial waiting time, only the
 states
 $\rho_{0000},\rho_{i000},\rho_{0i00},\rho_{00(ij)0},\rho_{000(ij)}$
 are significantly populated, while the population of all other
 states is small.
 In the next step it is shown that $\rho_M \approx \rho_{0000}$. In
 the following we will use the short-hand notation
 $\rho_0:=\rho_{0000}$.
 \subsubsection{Bounds for occupation probabilities}
 Let us define the probabilities $p_{ijkl}=\tr(\rho_{ijkl})$. A
 system of differential equations
 $\dot{p}_{ijkl}=\tr(\dot{\rho}_{ijkl})$ for these probabilities
 can be derived by from the differential equations for
 ${\rho}_{ijkl}$. Traceless terms such as $\gamma
 \mathfrak{L}(\rho)$ and $T_{ij}$ do no longer appear.

 We define $p_{000X},\dots,p_{XXX0},p_{XXXX}$ as the sum of
 $p_{ijkl}$, where all indices marked with $X$ are summed from $1$
 to $n,n'$, in order to remove the dependence on the Kraus
 operators by virtue of their normalization condition ($\sum
 _{i=1}^n A_i^\dagger A_i\!=\!\I$). A successful application of the
 LOCC map corresponds to the series $p_{X000}, p_{0X00}, p_{00X0},
 p_{000X}, p_{0000}$ (similarly to a X-excitation which created on
 the first index, travels to the right and disappears in the end).
 $p_{0000}$ takes high values, while the other probabilities are on
 the order of $\frac{\delta}{\Gamma}$, which indicates that this
 process is fast. However, the situation considered here does not
 correspond to this ideal case because Alice can start a new round
 before the last one is finished, which gives rise to probabilities
 which are denoted by indices with more than one $X$, e.g.
 $p_{XX00}, p_{X0X0},\dots$ which results in an incorrect
 realization of the LOCC map.
 Since we are not interested in the complete solution but only in
 upper and lower bounds, we further simplify the system by defining
 the probabilities
 $p_{0000},p_{X\Xi\Xi\Xi},p_{0X\Xi\Xi},p_{00X\Xi}$ and $p_{000X}$,
 which cover all possible events. $X$ indicates that the
 corresponding index is different from zero. Therefore, the
 summation runs from $1$ to $n,n'$. $\Xi$ stands for an arbitrary
 value, i.e., the summation starts from zero. These five quantities
 include also non-ideal process, with two or more X entries which
 correspond an errors and evolve according to
 \begin{eqnarray*}
 \dot{p}_{0000}&\!=\!&\! -\delta {p}_{0000} + \Gamma {p}_{000X},
 \\ \nonumber
 \dot{p}_{000X}&\!=\!& \!-(\delta+\Gamma) {p}_{000X} + \Gamma
 {p}_{00X\Xi},
 \\ \nonumber
 \dot{p}_{00X\Xi}&\!=\!& \!-(\delta+\Gamma) {p}_{00X\Xi} + \Gamma
 {p}_{0X\Xi\Xi},
 \\ \nonumber
 \dot{p}_{0X\Xi\Xi}&\!=\!& \!-(\delta+\Gamma) {p}_{0X\Xi\Xi} +
 \Gamma {p}_{X\Xi\Xi\Xi},
 \\ \nonumber
 \dot{p}_{X\Xi\Xi\Xi}&\!=\!& \!-\Gamma {p}_{X\Xi\Xi\Xi} \!+\!
 \delta\!( {p}_{0000}\!+\!
 {p}_{000X}\!+\!{p}_{00X\Xi}\!+\!{p}_{0X\Xi\Xi}).
 \end{eqnarray*}
 The solution shows that the steady state (ss) with
 \begin{eqnarray} \label{eq:ss}
 {p}_{0000}^{\text{ss}}\!\!&=&\!\!
 \frac{\Gamma^4}{(\Gamma\!+\!\delta)^4},\ \ \
 {p}_{000X}^{\text{ss}}\!\!=\!\!\frac{\delta
 \Gamma^3}{(\Gamma\!+\!\delta)^4}, \ \ \
 {p}_{00X\Xi}^{\text{ss}}\!\!=\!\! \frac{\delta
 \Gamma^2}{(\Gamma+\delta)^3},\nonumber
 \\
 {p}_{0X\Xi\Xi}^{\text{ss}}\!\!&=&\!\! \frac{\delta
 \Gamma}{(\Gamma\!+\!\delta)^2}, \ \
 {p}_{X\Xi\Xi\Xi}^{\text{ss}}\!\!=\!\!
 \frac{\delta}{(\Gamma\!+\!\delta)}.
 \end{eqnarray}
 is reached up to an error smaller than $\mathcal{O}(\alpha^2)$
 after a time of the order of $\frac{1}{\delta}$
 \cite{FootnoteSteadyStateError}. In the steady state,
 ${p}_{0000}^{\text{ss}}\!=\!1 \!-\! 4
 \frac{\delta}{\Gamma}\!+\!\mathcal{O}(\frac{\delta^2}{\Gamma^2})$.
 Next, bounds for ${p}_{X000},{p}_{0X00},{p}_{00X0},{p}_{000X}$ are
 derived. According to Eq.~(\ref{eq:x000}),
 $\dot{p}_{X000}=-\Gamma{p}_{X000}+\delta {p}_{0000}+\Gamma
 {p}_{X00X}$.
 Assuming that ${p}^{\text{ss}}_{0000}$ is reached after a time
 $t'$,
 \begin{eqnarray*}
 {p}_{X000}\!(t\!)\!\!\!&=&\!\!\!e^{-\Gamma(t\!-\!t'\!)}\!{p}_{X000}(t'\!)
  \!+\!\!\!\int_{t'}^t\!\!\!\! d\tau e^{\Gamma (\tau\!-\!t)}\!\!\left(\!\! \frac{\delta\Gamma^4}{( \Gamma\!+\!\delta)^4} \!+\! \Gamma
  {p}_{X00X}\!\!\right)\!\!,\\
  \!\!\!&=&\!\!\!e^{-\Gamma (t\!-\!t')}\!{p}_{X000}(\!t'\!)
  \!+\! (1\!-\!e^{-\Gamma (t\!-\!t'\!)}) \!\frac{\delta \Gamma^3}{(\Gamma\!\!+\!\!\delta)^4} +
  h(t,t'),
 \end{eqnarray*}
 where $h(t,t')\!\geq\!0$ is a positive function since
 ${p}_{X00X}\!\geq \!0$. Hence, doubling the initial waiting time
 guarantees ${p}_{X000}\!\geq\!\frac{\delta
 \Gamma^3}{(\Gamma\!+\!\delta)^4}$.
 According to Eq.~(\ref{eq:0x00}), $\dot{p}_{0X00}\!=\!-(\Gamma
 \!+\!\delta){p}_{0X00}\!+\!\Gamma{p}_{X000}\!+\!\Gamma
 p_{XX00}\!+\!\Gamma {p}_{0X0X}$. By integration, using the bound
 for ${p}_{X000}$ and assuming that the contributions from
 ${p}_{0X0X}$ and ${p}_{XX00}$ sum up to a positive function we
 conclude that $\frac{\delta
 \Gamma^4}{(\Gamma\!+\!\delta)^5}\!\leq\!{p}_{0X00}$ is fulfilled
 after waiting for another period on the order of
 $\frac{1}{\delta}$. Similarly, one obtains $\frac{\delta
 \Gamma^5}{(\Gamma\!+ \!\delta)^6}\!\leq\!{p}_{00X0}$ and $
 \frac{\delta \Gamma^6}{(\Gamma\!+\!\delta)^7}\!\leq\!{p}_{000X}$.
 Since
 $p_{0000}\!+\!p_{X000}\!+\!p_{0X00}\!+\!p_{00X0}\!+\!p_{000X}\!\geq\!1\!-\!
 \frac{12 \delta^2}{\Gamma^2}$, any probability with more than two
 $X$ entries is smaller than $\frac{12 \delta^2}{\Gamma^2}$. In
 summary,
 \begin{eqnarray*}
 \frac{\delta}{\Gamma}-7 \frac{\delta^2}{\Gamma^2}  \leq
 {p}_{X000},{p}_{0X00},{p}_{00X0},{p}_{000X} \leq
 \frac{\delta}{\Gamma},
 \end{eqnarray*}
 after a time of the order $\frac{1}{\delta}$, where the upper
 bounds are found using Eq.~(\ref{eq:ss}). Hence, a steady state is
 reached where states labelled with one (more than one) $X$ are
 occupied with probability $O(\alpha)$ ($O(\alpha^2)$).

 \subsubsection{Differential equation for $\rho_0$}
 The evolution of $\rho_0\equiv\rho_{0000}$ is governed by
 Eq.~(\ref{Eq:MEtotal}). After a period of the order
 $\frac{1}{\delta}$, $\dot{\rho}_{0}\!=\!\gamma'
 \mathcal{O}(1\!+\!\alpha)$ since
 $\|{\rho}_{000X}\|\!=\!\frac{\delta}{\Gamma}\!+\!\mathcal{O}(\alpha^2)$.
 Hence, for $\alpha\!\ll\!1$, $\rho_0$ is approximately constant on
 time scales that are short compared to $\gamma'$ .
 In order to obtain an equation which depends only on $\rho_0$, we
 solve successively the differential equations for
 ${\rho}_{i000},{\rho}_{0i00},{\rho}_{00(ij)0}$ and
 ${\rho}_{000(ij)}$. According to Eq.~(\ref{eq:x000}),
 \begin{eqnarray*}
  \dot{\rho}_{i000}
  \!\!\!&=&\!\!\!-\Gamma {\rho}_{i000} \!+\! \delta A_i {\rho}_{0}
 A_i^\dagger \!+\! \mathfrak{N},
 \end{eqnarray*}
 where  $\mathfrak{N}\!=\!\gamma
 \mathfrak{L}({\rho}_{i000})\!-\!\delta {\rho}_{i000}\!+\!\delta
 \sum_k A_i {\rho}_{k000} A_i^\dagger\!+\!\Gamma \sum_{xy=1}
 T_{xy}(\rho_{i00(xy)})$, which is bounded by $\gamma'
 \mathcal{O}(\alpha)$. For the first three terms we use
 $p_{X000}=\mathcal{O}(\alpha)$, the last term can be bounded by
 $p_{X00X}=\mathcal{O}(\alpha^2)$ such that
 \begin{eqnarray*}
 {\rho}_{i000}(t)\!=\!{\rho}_{i000}(0) e^{-\Gamma t} +\!\int_0^t
 \!d\tau e^{\Gamma(\tau-t)}\left(\delta  A_i {\rho}_{0}
 A_i^\dagger+\mathfrak{N} \right).
 \end{eqnarray*}
 The integral $\int_0^t d\tau e^{\Gamma(\tau-t)}\mathfrak{N} $ can
 be bounded by $\mathcal{O}(\alpha^2)$. The initial term is
 suppressed by $e^{-\Gamma t}$ and therefore smaller than
 $\mathcal{O}(\alpha^2)$ after the initial waiting time. Hence,
 \begin{eqnarray}\label{Eq:1}
 {\rho}_{i000}(t)&=& \mathcal{O}(\alpha^2) + \delta\int_0^t d\tau
 e^{\Gamma(\tau-t)}  A_i {\rho}_{0} A_i^\dagger.
 \end{eqnarray}
 Since the integral is mainly determined by terms close to $\tau
 \!=\!t$ and $\rho_0$ varies little on small time intervals,
 $\rho_0$ can be assumed to be constant. To prove this, we consider
  \bea \label{eq:23}
  \int_0^{t} X(\tau,t) d\tau =
 \int_0^{t'} X(\tau,t) d\tau + \int_{t'}^t X(\tau,t) d\tau ,\eea
 with $X(\tau,t)\!=\!e^{\Gamma(\tau\!-\!t)}  A_i {\rho}_{0}
 A_i^\dagger$ and $t'\!=\!t\!-\!\frac{1}{\sqrt{\Gamma
 \gamma'}}\!=\!t\!-\!\frac{1}{\gamma'} \sqrt{\alpha}$. Since
 $\frac{1}{\gamma'}$ is the typical time during which $\rho_0$
 changes, it is nearly constant during the interval $(t,t')$. From
 ${\dot \rho}_{0}=\mathcal{O}( \gamma' (1+\alpha))$, we obtain that
 for any $t'' \in [t',t]$
  \bea \label{eq:sqrtrho}
 \rho_{0}(t'')=\rho_{0}(t)+\int_t^{t''} d\tau \dot{\rho}_0(\tau)
 =\rho_{0}(t) +\mathcal{O}( \sqrt{\alpha}). \eea
 The integral from $0$ to $t'$ in Eq.~(\ref{eq:23}) are suppressed
 at least by a factor $e^{-\sqrt{\frac{1}{\alpha}}}<\alpha$ ($ A_i
 {\rho}_{0} A_i^\dagger$ is on the order of one). Inserting
 Eq.~(\ref{eq:sqrtrho}) in  Eq.~(\ref{Eq:1}) and using
 $\delta\int_{t'}^{t} d\tau
 e^{\Gamma(\tau-t)}=\frac{\delta}{\Gamma}(1-e^{-\sqrt{\alpha^{-1}}})$
 with $e^{-\sqrt{\alpha^{-1}}}<\alpha$ yields
  \bea \label{eq:sol1}
 {\rho}_{i000}(t)&=& \frac{\delta}{\Gamma} A_i {\rho}_{0}(t)
 A_i^\dagger + \frac{\delta}{\Gamma} \mathcal{O}(\sqrt{\alpha}),
 \\ \nonumber
 \eea
 which shows that for small $\alpha$, Alice applies her first POVM
 with high accuracy.

 Next, we consider the evolution of ${\rho}_{0i00}$ (see
 Eq.~(\ref{eq:0x00})),
 \bea \dot{\rho}_{0i00}=\ -\Gamma{\rho}_{0i00} +\Gamma
 \rho_{i000}+\mathfrak{N}, \eea
 where $\mathfrak{N}$ can be bounded by
 $\gamma'\mathcal{O}(\alpha)$ using
 $p_{0X00}\!=\!\mathcal{O}(\alpha)$,
 $p_{XX00}\!=\!\mathcal{O}(\alpha^2)$ and
 $p_{0X0X}\!=\!\mathcal{O}(\alpha^2)$. Inserting
 Eq.~(\ref{eq:sol1}) yields
 ${\rho}_{0i00}(t\!)\!\!\!=\!\!\!{\rho}_{0i00}(0) e^{-\Gamma
 t}\!\!+\! \! \!\int_0^t \!\!\!d\tau e^{\Gamma(\tau\!-\!t)}
 \!\!\left(\!\delta A_i {\rho}_{0}(\tau) A_i^\dagger\!+\!\delta
 \mathcal{O}( \sqrt{\alpha})\!+\!\mathfrak{N} \right)$.
 As before, the integral over $\mathfrak{N}$ and the first term can
 be bounded by $\mathcal{O}(\alpha^2)$ after a waiting time. If the
 remaining integral is split as in Eq.~(\ref{eq:23}), we obtain one
 part, where $\rho_0$ is nearly constant and one vanishing part.
 The main error is again due to expression (\ref{eq:sqrtrho}),
 leading to
 \bea {\rho}_{0i00}(t)&=&   \frac{\delta}{\Gamma} A_i {\rho}_{0}(t)
 A_i^\dagger +   \frac{\delta}{\Gamma} \mathcal{O}(\sqrt{\alpha}).
 \\ \nonumber
 \eea
 Hence, for small $\alpha$, sending classical information to Bob
 causes only marginal errors on the main system.

 Next, we consider the evolution of ${\rho}_{00(ij)00}$
 (Eq.~(\ref{eq:00x0})),
 \begin{eqnarray*}
  \dot{\rho}_{00(ij)0}= -\Gamma{\rho}_{00(ij)0}+\Gamma B^i_j\rho_{0i00}{B^i_j}^\dagger
 +\mathfrak{N},
 \end{eqnarray*}
 where $\mathfrak{N}$ can be bounded by $\gamma' O(\alpha)$ such
 that
 \bea {\rho}_{00(ij)0}(t)&=& \frac{\delta}{\Gamma} B^i_j A_i
 {\rho}_{0}(t) A_i^\dagger {B^i_j}^\dagger + \frac{\delta}{\Gamma}
 \mathcal{O}(\sqrt{\alpha}), \nonumber \eea
 which corresponds to a process, where Bob applies his part of the
 POVM and writes $i,j$ onto his classical input register.
 Similarly, Eq.~(\ref{eq:000x}) leads to
 \bea {\rho}_{000 (ij)}(t)&=& \frac{\delta}{\Gamma} B^i_j A_i
 {\rho}_{0}(t) A_i^\dagger {B^i_j}^\dagger + \frac{\delta}{\Gamma}
 \mathcal{O}(\sqrt{\alpha}),\nonumber \eea
 which corresponds to a transfer of the classical measurement
 results $i,j$ back onto Alice's side. Finally, these results can
 be applied for calculating $\rho_{0}$,
 \begin{eqnarray*}
 \dot{\rho}_{0}\!\!=\!\gamma \mathfrak{L}(\!\rho_{0}\!) \!
 \!-\!\delta {\rho}_{0} \!\!+\!\!\Gamma \!\sum_{ij}
 \!T_{ij}({\rho}_{000(ij)}\!) \!\!=\!\gamma
 \mathfrak{L}(\rho_{0})\!-\!\delta\!
 \left(T'(\!\rho_0\!)\!-\!{\rho}_{0}\!\right).
 \end{eqnarray*}
 $T'(\rho)\equiv T(\rho_0)+\mathcal{O}(\sqrt{\alpha})$ represents a
 noisy version of the desired LOCC map $T(\rho)$. The undesired
 contribution can be suppressed by choosing $\alpha$ small, i.e. by
 choosing $\Gamma$ large enough.
 The generalization to more than one round of communication is
 straight forward. By summing over the indices of the corresponding
 Kraus operators, one obtains equations for the probabilities,
 which are independent of the POVMs applied in the protocol. From
 these equations it can be concluded that only the relevant state
 responsible for the application of the LOCC is populated, while
 all others are suppressed by a factor of order $\alpha^2$ after  a
 initial waiting time. By successive integration as shown above,
 the desired approximation for $\dot \rho$ is obtained. Many
 distillation protocols only require only a small number of rounds
 to reach high fidelities and often even only one-way communication
 (half a round) [9,15].
\section{3. Stabilization of dissipative distillation schemes against errors acting on the target
system}
%
\begin{figure}
\includegraphics[width=8.5cm]{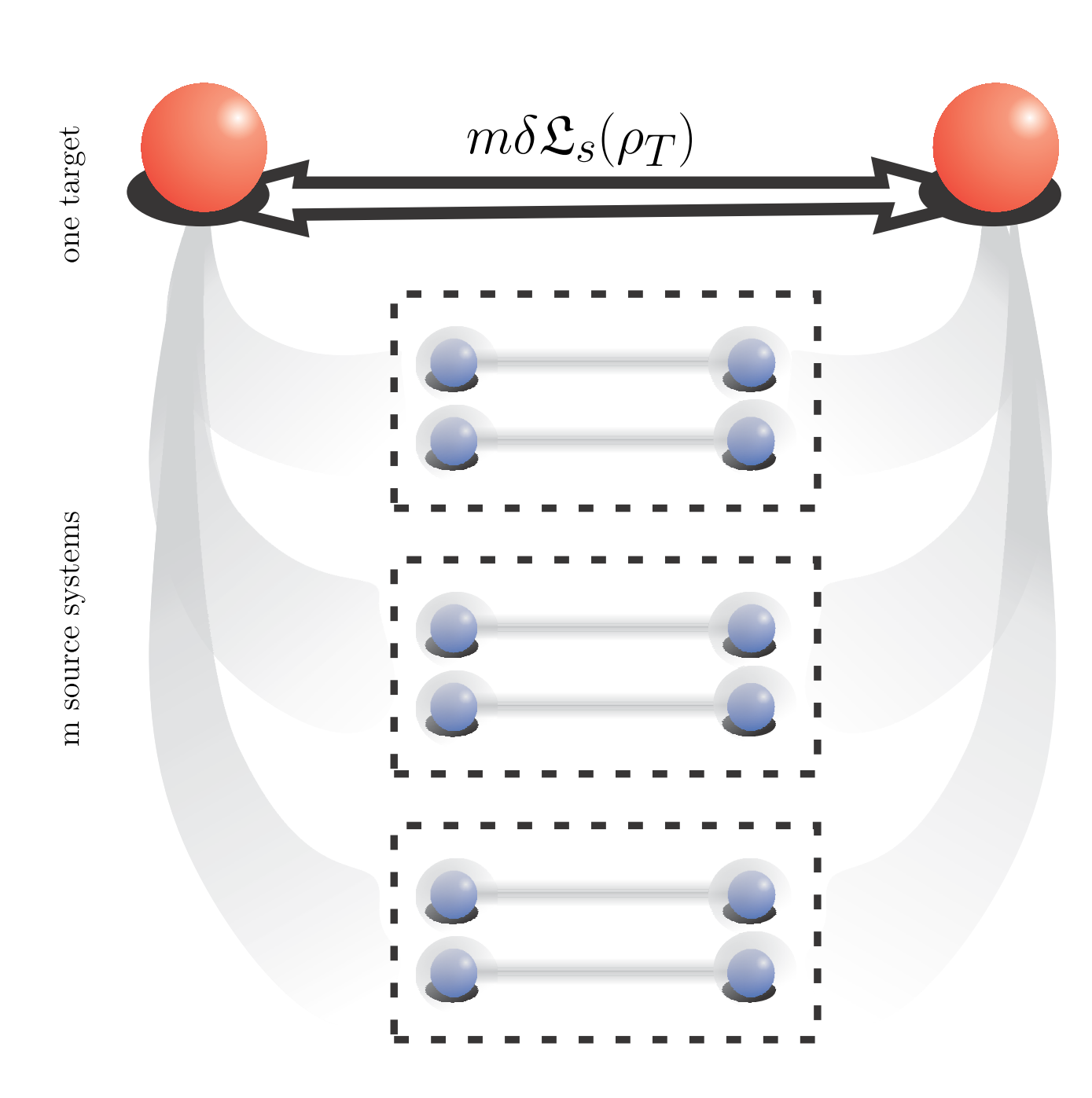}
\caption{(Color online) Stabilization of dissipative protocols
against noise acting on the target system by coupling several
source system to the same target.} \label{fig:booster}
\end{figure}

In this section, we explain how the distillation schemes presented
in Sec.~1.2 and Sec.~4 can be made robust against noise acting on
the target system. The same method for stabilization against
errors is applicable for both protocols and a wide range of other
dissipative schemes, which include classical communication.
The basic idea is illustrated in Fig.~\ref{fig:booster}. A
dissipative protocol is run using $m$ blocks of source systems in
parallel, which are all coupled individually to the same target
system. This way, a boost effect on the desired dynamics of the
target system can be achieved, while the back-action on the source
pairs remains unchanged. If sufficiently many source systems are
provided, the dynamics on the target system is dominated
completely by the desired dynamics.
In the following, we explain the application of this method first
for schemes of the type described in Sec.~4 and 5 and discuss then
briefly the stabilization of scheme I.
We start out by considering a target system $\T$ and a source
block consisting of $n$ pairs. An entangling dissipative process
described by the Lindblad operator $\delta \mathfrak{L}(\rho)$
acts on each source pair separately such that each of them is
individually driven into an entangled steady state $\rho_s$. The
effective master equation for the target system, which is obtained
by tracing out the source system, is given by
\begin{eqnarray*}
\dot{\rho}_\T=\delta \mathfrak{L}_S(\rho_\T),
\end{eqnarray*}
where the Lindblad operator $\mathfrak{L}_S(\rho_\T)$ may be time
dependent. It does not depend on the state of $\T$ but only on the
state of the source system as indicated by the subscript $S$.
Accordingly, the convergence speed at which
$\mathfrak{L}_S(\rho_\T)$ converges to a constant operator is
given by the rate at which the source system reaches a steady
state. The convergence rate of the source system is limited by the
rate $\delta$ at which the flip operation mapping the quantum
states of the source system to $\T$ is performed (see
Sec.~1 and Sec.~4).\\
\\We assume that $m$ identical source system $S_1, \dots, S_m$ are
individually coupled to a single source system $\T$ through the
Lindblad operator $\delta\sum_{i=1}^m\mathfrak{L}(\rho_{\T,S_i})$,
where $\mathfrak{L}(\rho_{T,S_i})$ is a Lindblad operator acting
on $\T$ and the $i$th source system (see Fig.~\ref{fig:booster}
for a schematic overview). We assume that these operators are
identical $\mathfrak{L}_{S_i}=\mathfrak{L}_{S}$ such that the
dynamics of the target system is governed by the reduced master
equation
\begin{eqnarray*}
\dot{\rho}_\T=\delta \sum_i \mathfrak{L}_{S_i}(\rho_\T)=m \delta
\mathfrak{L}_{S}(\rho_\T).
\end{eqnarray*}
This is not generally the case, since the $m$ source systems are
coupled to each other through the target system. Due to this
indirect coupling, the source systems may evolve differently in
time and can reach different steady states, which can be
disadvantageous for the evolution of the target system. This is
for example the case for the scheme described in Sec.~1.1 which
does not include classical communication.

It can be shown that
$\mathfrak{L}_{S_i}(\rho_\T)=\mathfrak{L}_{S}(\rho_\T)$, if there
is no state dependent back-action of $\T$ on the source systems.
In this case, the evolution of the reduced density matrix of each
source block is independent from the time evolution of the other
blocks. This property can be guaranteed by re-initializing the
source systems after each swap operation in a standard state, for
example the identity (strict equality requires in principle also
that all source systems start from the same initial state.
However, different initial states have only an effect on the time
evolution in the beginning. The following discussions are only
concerned with the steady state of the system, which is
independent of the initial conditions).
Scheme I including classical communication (see Sec.~1.2) exhibits
a weak state dependent back-action. As explained in the end of
Sec.~1.2, this can be avoided by applying a twirl
\cite{HWernerStates} on the target system prior to each flip
operation. Hence, the stabilization method outlined above is
directly applicable to this modified version of the scheme
\cite{Footnote3}.\\
\\By boosting the desired dynamics on the target system, arbitrary
high error rates $\epsilon$ can be tolerated. For $m \delta \gg
\eps$, the dynamics governed by the master equation
\begin{eqnarray*}
\dot{\rho}_T=m \delta  \mathfrak{L}_s(\rho_T) + \eps
\mathfrak{L}_{\text{noise}}(\rho_T)
\end{eqnarray*}
is dominated by the first term and the steady state is arbitrarily
close to the original steady state.
In the specific case, where the process $\mathfrak{L}_s(\rho_\T)
=\tr{(\rho_\T)}\rho_{\T\!,s}-\rho$ driving the target system into
the steady state $\rho_{\T\!,s}$ is counteracted by depolarizing
noise $(\tr{(\rho_\T)} \1-\rho_\T)$, the time evolution described
by
\begin{eqnarray*}
\dot{\rho}_\T=m \delta (\tr{(\rho_\T)}\rho_{\T\!,s}-\rho) + \eps
(\tr{(\rho_\T)} \1-\rho_\T)
\end{eqnarray*}
leads to the steady state $ \rho_{\T,\!s}'=\frac{m \delta
\rho_{\T\!,s}+ \eps \1}{m \delta +\eps}$, which can be easily
verified by solving the equation $\dot{\rho}_\T=0$. This state is
reached exponentially fast with a rate $m \delta+\eps$.
The same result holds for local depolarizing noise acting on
Alice's and Bob's system (see Sec.~4.1) if the steady state
$\rho_{\T\!,s}$ is a Werner state. A master equation of this type
is solved exactly in the next section.
\section{4. Scheme II: dissipative entanglement distillation for Werner states}
In this section, we introduce a second dissipative distillation
scheme, which does not rely on entangling processes producing
steady states, which are close to pure states, as scheme I
presented in Sec.~1.
We analyze here a very general model for Werner states
\cite{HWernerStates}, which can be solved exactly. Werner states
are of the simple form $\rho_{\text{W}}(f)=f \Omega
+(1-f)(\I-\Omega)/3$, and are characterized in terms of their
fidelity $f$, which is given by the overlap with the maximally
entangled state $\Omega$.
Any quantum state can be transformed into a Werner state by
twirling \cite{HWernerStates} without a loss of fidelity. Since a
Werner-twirl is a LOCC map, a dissipative protocol can be
constructed, which corresponds to the continuous application of a
twirl operation on a given system and mapping of the resulting
state to a new pair acting as target system $\T$ by means of a
continuous flip procedure (compare Sec.~4.2).

This way, any dissipative process can be modified such that it can
be described in terms of a Werner Lindblad operator $E_f(\rho_\T)$
as used in Secs.~4 and 5, where $f$ is the steady state fidelity
of the underlying process. In this sense, the Werner model used
here is very general and can be applied in many situations.

\subsection {4.1 Dissipative entangling model process for a single source pair}
A dissipative model process, which produces an arbitrary Werner
state as steady state can be modelled by considering two
processes, which generate the steady states $\Omega$ and $\1$
respectively, where $\1=\I/4$ denotes the normalized identity. Let
$\ket{\psi_i}$ denote the four Bell-states, where
$\ket{\psi_0}=\left(\ket{00}+\ket{11}\right)/\sqrt{2}$, and
$\sigma_i$ the Pauli matrices, where $\sigma_0$ is the identity. A
master equation which leads to the steady state
$\Omega=\ket{\Psi_0}\bra{\Psi_0}$ can be constructed using the
four jump operators $Q_i=\ket{\psi_0}\bra{\psi_i}$,
which give rise to the Lindlbad term
\begin{eqnarray*}
\nonumber Q(\rho)=\sum_i
\mathfrak{L}^{Q_i}(\rho)=\text{tr}(\rho)\Omega-\rho.
\end{eqnarray*}
Similarly, a master equation which leads to the steady state $\1$
is obtained using the jump operators $ W_{ij}=\sigma_i \ot
\sigma_j$, which give rise to the Lindblad term
\begin{eqnarray*}
W(\rho)=\sum_{ij}
\mathfrak{L}^{W_{ij}}(\rho)=\text{tr}(\rho)\1-\rho.
\end{eqnarray*}
Hence, the Werner state $\rho_W(f)$ with fidelity $f$ is the
steady state of the time evolution governed by the master equation
\begin{eqnarray*}
E_f(\rho)= f Q(\rho)+ \frac{1-f}{3}(3
W(\rho)-Q(\rho))=\text{tr}(\rho)\rho_{W}-\rho.
\end{eqnarray*}
The Lindblad term $E_f(\rho)$ will be used in the following to
model the basic entangling process acting on the source systems.\\
\\Local depolarizing noise acting on Alice's (Bob's) side is
included using the jump operators $S_i=\I_A\otimes \sigma_i$
($S_i=\sigma_i\otimes\I_B$),
 such that the corresponding Lindblad
terms are given by
\begin{eqnarray*}
N_{\text{\tiny{Alice}}}(\rho)&=&\sum_i
\mathfrak{L}_{\text{\tiny{Alice}}}^{S_i}=\rho_\text{A}\otimes
\1_B-\rho,\\
N_{\text{\tiny{Bob}}}(\rho)&=&\sum_i
\mathfrak{L}_{\text{\tiny{Bob}}}^{S_i}=\1_A\otimes\rho_\text{B}-\rho,
\end{eqnarray*}
where $\rho_{\text{A}}$ ($\rho_{\text{B}}$) is the reduced density
matrix corresponding to Alice's (Bob's) system and $\1_A$ ($\1_B$)
the normalized identity $\I/2$ on  Alice's (Bob's) system. This
process describes the continuous replacement of the state on
Alice's (Bob's) side by the completely mixed state. The total
master equation
\begin{eqnarray}
\dot{\rho}=\gamma E_f(\rho)+ \frac{\eps}{2}
N(\rho)\label{Eq:ME:EntanglingProcess}
\end{eqnarray}
where
$N(\rho)=N_{\text{\tiny{Alice}}}(\rho)+N_{\text{\tiny{Bob}}}(\rho)$,
describes the basic entangling process including local noise. This
type of equation will be used frequently in the following
sections, as it also describes also the evolution of the target
systems once the corresponding source systems have reached the
steady state.\\
\\The steady state of the time evolution described by
Eq.~(\ref{Eq:ME:EntanglingProcess}) is a Werner state
\begin{eqnarray}
\rho_s=\frac{  \gamma \rho_{\text{W}}(f) +\eps
\1}{\gamma+\eps}\label{Eq:SS:EntanglingProcess}
\end{eqnarray}
with reduced fidelity $f_s=\frac{ \gamma f +\eps
\frac{1}{4}}{\gamma+\eps}$.
The general time dependent solution of the master equation
(\ref{Eq:ME:EntanglingProcess}) is of the form
\begin{eqnarray}
\rho(t)=\rho_0 g_0(t)+\rho_1 g_1(t)+\rho_2 g_2(t)+\rho_3
g_3(t),\label{los}
\end{eqnarray}
where $\rho_0$ is any initial state,
$\rho_1=\frac{1}{2}(\rho_{0,\text{A}} \otimes \1_B +\1_A \otimes
\rho_{0,\text{B}})$, $\rho_2=\rho_{\text{W}}(f)$ and $\rho_3
=\1=\1_A \otimes \1_B$. $\rho_{0,\text{A}}$ and $\rho_
{0,\text{B}}$ are the reduced density matrices of the initial
state $\rho_0$ at Alices and Bobs side.
The functions $g_i$ are given by
\begin{eqnarray}\label{eq:solu}
g_0(t)&=&e^{-\gamma' t}\nonumber,\\
g_1(t)&=&2\left( e ^{-\gamma'' t}-e ^{-\gamma' t}
\right),\\\nonumber
g_2(t)&=&\frac{\gamma}{\gamma'}(1-e^{-\gamma't}),\\\nonumber
g_3(t)&=&\frac{2 \gamma}{\gamma'} \left(e^{-\gamma' t}-e^{\gamma
'' t} \right)+\frac{\eps}{\gamma'}(e^{\gamma 't }-2 e^{\gamma ''t
} +1 ),
\end{eqnarray}
where $\gamma'=\gamma+\eps$ and $\gamma''=\gamma+\eps/2$. Note,
that the terms which depend on the initial state of the system,
i.e. $\rho_0$ and $\rho_1$, are suppressed exponentially fast. The
system reaches the steady state given by
Eq.~(\ref{Eq:SS:EntanglingProcess}) exponentially fast with a rate
of at least $\gamma+\frac{\eps}{2}$.\\

In order to verify that Eqs.~(\ref{los}) and~(\ref{eq:solu}) are a
solution of Eq.~(\ref{Eq:ME:EntanglingProcess}), Eq.~(\ref{los})
can be used as ansatz. The master equation gives rise to a set of
differential equations for the functions $g_i$ with initial
conditions $g_0=1$ and $g_i=0$,
\begin{eqnarray}
\dot g_0&=&-(\gamma+\eps) g_0,\\\nonumber \dot
g_1&=&-(\gamma+\frac{\eps}{2})g_1    +\eps g_0,
\\\nonumber \dot g_2&=&-\eps f_2 +\gamma
(g_0+g_1+g_2+g_3),\\\nonumber \dot g_3&=&-\gamma g_3 +
\frac{\eps}{2}g_1+ \eps g_2.
\end{eqnarray}
Below, the initial condition $\rho_0=\1$ will be considered
frequently. In this case the solution simplifies to
\begin{eqnarray}
\label{eq:start1} \rho(t)=\rho_s+(\1-\rho_s) e^{-(\gamma+\eps)t}.
\end{eqnarray}

\subsection {4.2 Steady state entanglement distillation acting on $n$ source systems}
We consider $n$ systems which are subject to the basic entangling
process $\gamma E_f(\rho)+\frac{\eps}{2} N(\rho)$ and are driven
into the steady state $\rho_s$ as described in Sec.~4.1.
These qubit pairs act as source systems for a LOCC distillation
operation $T_D$, which distills one potentially higher entangled
state from these copies. The resulting quantum state is mapped to
a target pair $\mathcal{T}$ and each source system is
re-initialized in the state $\1$. We do not specify $T_D$ at this
point - the solution derived in this section holds for any $n$ to
$1$ distillation protocol. We start out by considering only
deterministic protocols and generalize the results at the end of
this section such that probabilistic schemes are also covered.
Note that the complete re-initialization of the source systems
 represents the worst-case situation regarding the back-action
of the target system onto the source pairs. This choice allows us
solve the model exactly and to provide a lower bound for
dissipative distillation schemes of this type.\\
\\The continuous distillation procedure explained above is
described by the master equation
\begin{eqnarray}
\dot{\rho}= \sum_{i=1}^n  \left( \gamma E_f(\rho)+\frac{\eps}{2}
N(\rho) \right)_i +\delta_{D}(T_D(\rho)-\rho),
\end{eqnarray}
where $\left(X(\rho)\right)_i$ stands for the dissipative process
$X(\rho)$ acting on the $i$th source system.\\
\\In the following, we determine the time evolution and the steady
state of the target system. The reduced master equation for $
\mathcal{T}$ depends on the steady state of the reduced source
system. Therefore, we start by solving the dynamics on the source
system. Since the back-action of $ \mathcal{T}$ on the source
system does not depend on the quantum state of $ \mathcal{T}$, the
time evolution of the source pairs can be considered independently
from the target system.\\
For clarity, the reduced states of source and target system are
denoted by $\sigma$ and $\rho_\T$ respectively in this section.
The reduced master equation for the $n$ source systems is given by
\begin{eqnarray}
\dot{\sigma}= \sum_{i=1}^n \left( \gamma
E_f(\sigma)+\frac{\eps}{2} N(\sigma) \right)_i +\delta_{D}
(\tr{(\sigma)}\  \1^{\ot n}-\sigma). \label{mastersigma}
\end{eqnarray}
The solution of the homogeneous master equation which describes
the entangling dynamics for $n$ independent source systems
\begin{eqnarray*}
\dot{\sigma}_*\left(\sigma_0,t\right)=\sum_{i=1}^n \left(\gamma
E_f(\sigma_*(\sigma_0,t))+\eps N(\sigma_*(\sigma_0,t)) \right)_i,
\end{eqnarray*}
is already known (see Sec.~4.1) if the initial state is a product
state. $\sigma_*(\sigma_0,t)$ denotes the solution of the
homogeneous master equation with initial condition
$\sigma_*(\sigma_0,t=0)=\sigma_0$.
The solution of the inhomogeneous master equation
Eq.~(\ref{mastersigma}) is given by
\begin{eqnarray*}
\sigma(t)&=&\sigma_*(\sigma_0,t) e^{-\delta_D t}+\delta_D \int_0^t
d\tau \sigma_*(\1^{\ot n},t-\tau) e^{-\delta_D(t-\tau)},
\\ \nonumber
&=&\sigma_*(\sigma_0,t) e^{-\delta_D t}+ \delta_D \int_0^t d\tau
\sigma_*(\1^{\ot n},\tau) e^{-\tau \delta_D},
\end{eqnarray*}
with arbitrary initial condition $\sigma(0)=\sigma_0$. This
solution can be easily verified by considering the time derivative
\begin{eqnarray*}
\dot{\sigma}(t)&=&-\delta_D \sigma(t)+e^{-\delta_D
t}\dot{\sigma}_*(\sigma_0,t) \\ \nonumber &+& e^{-\delta_D
t}\partial_t\left[ \delta_D \int_0^t d\tau \sigma_*(\1^{\ot
n},t-\tau) e^{\tau \delta_D} \right].
\end{eqnarray*}
Using $\partial_t\int_0^t g(\tau)f(t-\tau)=f(0)g(t)+\int_0^t
g(\tau) \dot{f}(t-\tau)$ and $\sigma_*(\1^{\ot n},0)=\1^{\ot n}$,
one obtains
\begin{eqnarray*}
\dot{\sigma}(t)&=&-\delta_D \sigma(t)+e^{-\delta_D
t}\dot{\sigma}_*(\sigma_0,t) \\ \nonumber &+& \delta_D{\1^{\ot
n}}+e^{-\delta_D t}\partial_t\left[ \delta_D \int_0^t d\tau
\dot{\sigma}_*(\1^{\ot n},t-\tau) e^{\tau \delta_D} \right],
\end{eqnarray*}
which yields Eq.~(\ref{mastersigma}). The steady state
\begin{eqnarray}\label{Eq:SS:sigma}
\sigma_s=  \delta_D \int_0^\infty d\tau \sigma_*(\1^{\ot n},\tau)
e^{-\delta_D  \tau}
\end{eqnarray}
is reached exponentially fast with a rate of at least $\delta_D$.
The homogeneous solution $\sigma_*(\1^{\ot n},\tau)$ is given by
the tensor product of the solution for a single source pair
(\ref{eq:start1}),
\begin{eqnarray*}
\sigma_*(\1^{\ot n},t)=(\rho_s+(\1-\rho_s) e^{-(\gamma+\eps)t}
)^{\ot n},
\end{eqnarray*}
such that Eq.~(\ref{Eq:SS:sigma}) can be further simplified
\begin{eqnarray}
\sigma_s=\int_0^1 dx (\rho_s +(\1-\rho_s)
x^{\frac{\gamma+\eps}{\delta_D}})^{\ot
n}.\label{Eq:SteadyStateIntegralx}
\end{eqnarray}
Next, we consider the dynamics of the target system $\mathcal{T}$
described by the time dependent master equation
\begin{eqnarray*}
\dot{\rho}_{\mathcal{T}}=\delta_{D}
\left(T_D(\sigma(t))-\rho_{\mathcal{T}} \right),
\end{eqnarray*}
which is solved by
\begin{eqnarray*}
\rho_{\mathcal{T}}(t)=\rho_{\mathcal{T}}(0) e^{- \delta_D t}
+\int_0^t  d\tau \delta_D T_D(\sigma(t)) e^{- \delta_D(t-\tau)}
\end{eqnarray*}
with steady state $T_D(\sigma_s)$. The corresponding steady state
fidelity can be inferred by integrating over the fidelities that
are obtained if a standard distillation protocol is applied such
that
\begin{eqnarray*}
f_{\text{out}}(f_s)\!\equiv\! f_{\text{out}}(f,\eps)\!=\!\int_0^1
\! dx f_D(f_s \!+\!(\frac{1}{4}\!-\!f_s)
x^{\frac{\gamma+\eps}{\delta_D}})^{\ot n},
\end{eqnarray*}
where Eq.~(\ref{Eq:SteadyStateIntegralx}) was used.\\
\\So far, it has been assumed, that the underlying distillation protocol
$T_D$ is deterministic, such that a distilled state is available
whenever it is applied. However, many distillation protocols of
interest are probabilistic, i.e., they only succeed some
probability $P(\rho)$.
If a probabilistic distillation protocol is used, the
corresponding map $T_D$ is defined in such a way, that a flip
operation is only performed when the distillation was successful,
which leads to a state dependent rate in the master equation
\begin{eqnarray*}
\dot{\rho}_T=\delta_{D}P(\sigma(t)) \left(T_D(\sigma(t))-\rho_T
\right).
\end{eqnarray*}
Accordingly, the target system is driven into the same steady
state as discussed above with a reduced rate. Once the time
evolution of the source system has reached a steady state, the
dynamics of the target system is determined by the master equation
\begin{eqnarray*}
\dot{\rho}_T=\delta_{D} P
(\sigma_s)\left(\tr(\rho_T)\rho'_s-\rho_T \right)=\delta_D p
E_{f_{\text{out}}}(\rho_T),
\end{eqnarray*}
where $\rho'_s$ is the distilled steady state of the source
system. Since $\rho'_s$ is a Werner state, the target system can
act as one of $n$ new source systems which drive a new target
system into an even more entangled state. This way, the
distillation protocol can be iterated in a nested form.

\section{5. Continuous quantum repeaters}
The ability to distribute entangled states of high quality over
long distances is of vital importance for quantum communication
and quantum network related applications in general. As opposed to
classical information, quantum information cannot be cloned.
Therefore, classical repeater schemes are not applicable in this
context and quantum repeater schemes which respect the coherence
of quantum states are required
\cite{Briegel,DLCZ,QuantumInternet}. In quantum repeater
protocols, entanglement is first distributed over short distances
$L_0$ with high accuracy. Then neighboring pairs are connected by
a teleportation procedure \cite{Teleportation} (entanglement
swapping \cite{EntaglementSwappingTh,EntanglementSwappingExp})
such that entangled links which span a distance $2 L_0$ are
obtained. In the next step, two neighboring links of length $2
L_0$ are connected by entanglement swapping, resulting in
entangled pairs which span a distance $4L_0$. This way, an
entangled link of length $L=L_02^k$ can be established in $k$
iteration steps (compare Fig.~4 in the main text). However, for
non-maximally entangled states, entanglement swapping leads to a
considerable degradation in the fidelity of the resulting quantum
state. Since the distributed entanglement decreases dramatically
every time the length of the entangling links is doubled, it can
not be distributed over large distances this way. Therefore an
entanglement distillation protocol has to be applied after every
entanglement swapping procedure before proceeding to the next
stage.\\
\\In the following, we describe a continuous dissipative quantum
repeater scheme, which combines continuous swap and distillation
processes in order to generate long-range entangled steady states,
while entangling dissipative processes are only required over
short distances. To this end, we introduce a continuous swap
operation in Sec.~5.1 and explain in Sec.~5.2 how this method can
be combined with the distillation scheme presented above (Sec.~4)
such that a high-quality entangled link can be established over a
large distance as steady state of a continuous dissipative
evolution. We conclude this proof-of-principle study by giving a
specific example.
\subsection{5.1 Continuous entanglement swapping}\label{SubSec:Swapping}
%
\begin{figure}
\includegraphics[width=8.5cm]{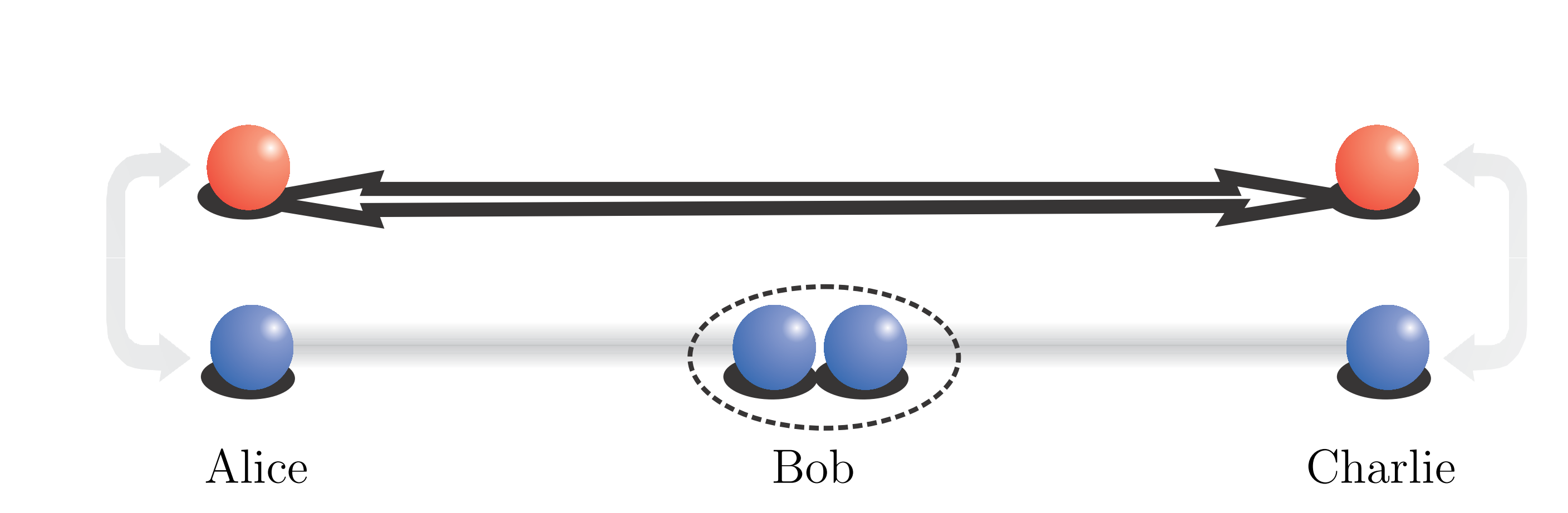}
\caption{(Color online) Continuous entanglement swapping.}
\end{figure}
%
The basic setup for entanglement swapping consists of three nodes
aligned on a line, operated by  Alice, Bob and Charlie, where
Alice and Bob as well as Bob and Charlie share an entangled pair,
while the distance between Alice and Charlie is too large for
generating an entangled state of high quality (see Fig.~3b in the
main text). By performing a teleportation procedure, which
requires the measurement of the two qubits at Bob's node and
classical communication to Alice and Charlie, as well as local
operations on their sides, an entangled link can be established
between Alice and Charlie \cite{EntaglementSwappingTh}.\\
\\We consider a setting, where Alice and Bob as well as Bob and
Charlie each hold a source pair which is subject to the basic
dissipative entangling mechanism considered in Sec.~4, such both
pairs are individually driven into the steady state $\rho_s$. This
dynamics is described by the Lindblad term
$\gamma_{\text{sw}}\sum_{i=1}^2 \left( E_f(\rho)
\right)_i=\gamma_{\text{sw}}\sum_{i=1}^2(\tr(\rho)
\rho_W-\rho)_i$. As illustrated in Fig.~3b in the main text, the
source pairs are coupled to a pair of target qubits shared between
Alice and Charlie through the term
$\delta_{\text{sw}}\left(T_{\text{sw}}(\rho)-\rho\right)$, where
the completely positive map $T_{\text{sw}}$ corresponds to a flip
operation which maps the state resulting from the entanglement
swapping procedure to a target system and re-initializes the
source systems in the state $\1 \ot \1$ \cite{FootnoteSwapping}.
Hence, the total master equation is given by
\begin{eqnarray*}
\dot\rho=\gamma_{\text{sw}} \sum_{i=1}^{2}  \left(
E_f(\rho)+\frac{\eps}{2} N(\rho)
\right)_i+\delta_{\text{sw}}\left(T_{\text{sw}}(\rho)-\rho\right)
\end{eqnarray*}
and the reduction to the source systems $\sigma$ yields
\begin{eqnarray*}
\dot\sigma=\gamma_{\text{sw}} \sum_{i=1}^{2} \left(
E_f(\rho)+\frac{\eps}{2} N(\rho) \right)_i
+\delta_{\text{sw}}\left(\tr(\sigma) \1 \ot \1-\sigma\right).
\end{eqnarray*}
The solution of this differential equation (compare Sec.~4.2)
\begin{eqnarray*}
\sigma(t)\!=\!\sigma_*\!(\sigma_0,\!t) e^{-\delta_{\text{sw}}
t}\!+\!\delta_{\text{sw}}\!\!\int_0^t \!\!d\tau \sigma_*\!(\1^{\ot
2},\!t\!-\!\tau) e^{-\delta_{\text{sw}} (t\!-\!\tau\!)},
\end{eqnarray*}
where $\sigma_*\!(\sigma_0,\!t)$ is the homogeneous solution with
initial condition $\sigma(t=0)=\sigma_{0}$, shows that the steady
state
\begin{eqnarray}
\sigma_s=\int_0^1 dx (\rho_s +(\1-\rho_s)
x^{\frac{\gamma_{\text{sw}}+\eps}{{\delta_\text{sw}}}})^{\ot
2},\label{Eq:Solution:Swap}
\end{eqnarray}
where $\rho_s=\frac{\gamma_{\text{sw}} \rho_W(f) +\eps
\1}{\gamma_{\text{sw}}+\eps}$, is reached exponentially fast with
a rate of at least $\delta_{\text{sw}}$.
\begin{figure*}
\includegraphics[width=16.5cm]{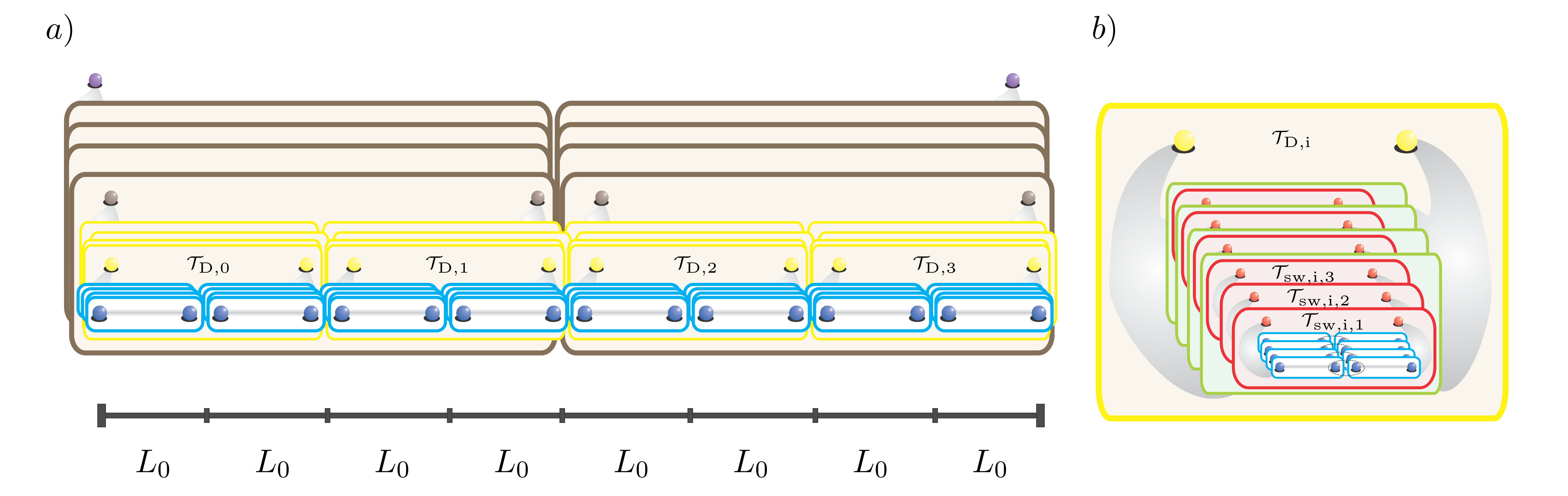}
\caption{(Color online) Dissipative quantum repeater architecture.
a) Concatenation of elementary steps, in which the distance over
which entanglement is distributed is doubled. b) Illustration of a
single iteration step including entanglement swapping and
-distillation.}\label{Fig:Repeatera}
\end{figure*}
\\The time dependent master equation governing the dynamics of the
target system is given by
\bea \label{eq:master} \dot{\rho}=\delta_{\text{sw}}
(T_{\text{sw}}(\sigma(t))-\rho_T),
 \eea
where $T_{\text{sw}}(\sigma_s)$ is the steady state of this
evolution. According to Eq.~(\ref{Eq:Solution:Swap}), the steady
state fidelity is given by
\begin{eqnarray*}
f_{\text{sw}}(f,\eps)\equiv f_{\text{sw}}(f_s)=\int_0^1 dx
F\left(f_s +(\frac{1}{4}-f_s)
x^{\frac{\gamma_{\text{sw}}+\eps}{{\delta_\text{sw}}}}\right),
\end{eqnarray*}
where $f_s$ is the fidelity of the state $\rho_s$ and
$F(f)=\left(1-2f+4f^2\right)/3$ is the output fidelity of the swap
protocol for two input states with fidelity $f$. A short
calculation shows that
\begin{eqnarray}
f_{\text{sw}}(f,\eps)\!\equiv\!f_{\text{sw}}(f_s)\!=\!2
\gamma_{\text{sw}}^2\frac{\left(\!\frac{1}{3}\!\left(1\!-\!2 f_s
\!+\!4 f_s^2\right)\!-\!\frac{1}{4}\right)\!+\!\frac{1}{4}}{(2
\gamma_{\text{sw}}+\delta_{\text{sw}})(\gamma_{\text{sw}}+\delta_{\text{sw}})}\
,\label{Eq:FidelitySwapping}
\end{eqnarray}
where $f$ is the fidelity of the state $\rho_W$ and
$f_s=\frac{\gamma_{\text{sw}} f+\eps
\frac{1}{4}}{\gamma_{\text{sw}}+\eps}$.
As discussed in Sec.~3, the scheme can be made robust against
noise processes acting on the target system by using $m$ copies of
the source systems and coupling them all to the same target state.
\subsection{5.2 Creation of long-range, high-quality steady state entanglement}\label{SubSec_Repeater}
The continuous swap operation introduced above (Sec.~5.1), the
dissipative distillation protocol explained in Sec.~4 and the
method for stabilization against errors acting on target systems
(Sec.~3) are the basic building blocks for the dissipative quantum
repeater scheme illustrated in Fig.~\ref{Fig:Repeatera}.
To begin with, the distance $L$ over which an entangled link has
to be established is divided into $2^k$ segments of length $L_0$,
as in standard repeater schemes. At each intermediate node, many
qubits are supplied which are subject to local depolarizing noise
acting at a rate $\eps$. We assume that each source pair
constituting an elementary link of length $L_0$ is individually
driven into a steady state $\rho_s$ of high fidelity by means of
an entangling dissipative process of the type discussed in
Sec.~4.1., $\gamma E_{f_I}(\rho)$, with high initial steady state
fidelity $f_I$ and a rate $\gamma$, which is large compared to the
noise rate $\epsilon$. Note that this assumption can also be
satisfied starting from dissipative processes leading to a steady
state with low fidelity and low $\gamma$ if distillation and boost
processes are applied as discussed above. In the following, we
consider an iteration step of the repeater protocol which acts on
$2^r$ entangled source systems, which each span a distance $l$
with fidelity $f_l$, and produces entangled links of the length
$2l$ with fidelity $f_{2l}$, such that $f_{2l}\geq f_{l}$. This is
illustrated in Fig.~\ref{Fig:Repeatera}, where the entangled
source pairs of length $l$ are shown in blue and the yellow target
pairs of length $2l$ are depicted in yellow. Each iteration step
consist out of the following subroutines, which are illustrated in
Fig.~\ref{Fig:Repeatera}b:
\begin{itemize}
  \item Neighboring source pairs of length $l$ (blue) are connected via a
        continuous swap operation. The resulting quantum states are
        written onto target pairs $\mathcal{T}_{\text{sw},i}$
        (red). In order to achieve a boost-effect on the targets, this protocol is run on $m$ source systems in
        parallel.
  \item A block of $n$ such pairs $\mathcal{T}_{\text{sw},i,j}$ , $j=1,\dots,n$, acts as source system (green) for an distillation process, which maps the resulting quantum state to new target system $\mathcal{T}_{\text{D},i}$ (yellow).
  \item $m$ of these blocks (green) are needed to achieve a high fidelity of the quantum state of the target systems (yellow).
\end{itemize}
This iteration step results in entangled links (yellow) which span
twice the initial distance and feature a high fidelity as well as
a high convergence rate once all source systems have reached the
steady state.\\
\\In the following, we consider
$\delta_{D}=\delta_{sw}=\frac{\gamma}{m}$ for simplicity (these
parameters can be optimized for a given distillation protocol).
The individual levels of the repeater scheme converge seriatim
from bottom to top to a steady state. For example, once the source
systems of length $l$ (blue) are in a steady state, the reduced
master equation for the target system of length $2l$ (yellow)
becomes time independent and this system reaches a steady state
too.
We assume now, that all source pairs of length $l$ (blue) are
driven by a time independent master equation of the type discussed
in Sec.~4.2, $\dot{\rho}=\gamma E_{f}(\rho)$ once all underlying
systems have reached the steady state.
The reduced master equation for the target system
$\mathcal{T}_{\text{sw},i}$ (red)
\begin{eqnarray} \nonumber
\dot{\rho}_{\mathcal{T}_{\text{sw}},i}\!=\!\delta m
E_{f_{sw}(f_I,\eps)}({\rho}_{\mathcal{T}_{\text{sw}},i})
\!+\!\frac{\eps}{2}N({\rho}_{\mathcal{T}_{\text{sw}},i})
\!+\!\delta \left(\1^{\ot n}\!-\!\rho_{\mathcal{T}_{\text{sw}},i}
\right )\!
\end{eqnarray}
includes local polarizing noise as introduced in Sec.~4.1 and the
back-action of the distillation scheme. Note, that the rate of the
entangling process, $\delta m= \gamma$ is again high, due to the
boost on the target system. The entangling process acting on the
target systems of the distillation procedure $\T_{D},i$ (yellow),
$E_{f_{\mathcal{T}_{\text{sw}}}}(\rho_\mathcal{T_{\text{D}}} )$ is
determined by the steady state fidelity
$f_{\mathcal{T}_{\text{sw}}}:=f_{sw}(f_I,\eps)$.\\
A wide range distillation protocols for Werner states
\cite{protocols} can be used in a continuous form as demonstrated
in Sec.~4 (below a specific example is discussed). As explained
there, a distillation protocol corresponds to a completely
positive map $T_{D}$ which is described by a Linblad term
$\delta_{\text{D}}\left(T_{D}(\rho)-\rho\right)$. The distillation
process is applied continuously for each entangled link and the
resulting highly entangled qubit state is flipped to new target
pairs $\mathcal{T}_{\text{D},i}$ spanning the same length $l$.
\\We consider here a distillation procedure which acts on $n$
entangled source systems and distills one potentially higher
entangled pair. Hence, for each of the $2^{r-1}$ links, $n$ copies
$\mathcal{T}_{\text{sw},ij}$, $i\in\{1,2^{r-1}\}$, $j\in\{1,n\}$
have to be supplied. This situation is sketched in
Fig.~\ref{Fig:Repeatera}b, where the target systems
$\mathcal{T}_{\text{sw},ij}$ (red), driven by the source pairs
(blue), are used as resource for creating a highly entangled
steady state of the new target pair (yellow). One source block
(shown in green) is sufficient for entanglement distillation, but
several of them running in parallel are needed to boost the
desired dynamics on the target system.
This way, each target pair $\mathcal{T}_{\text{D},i}$ is driven at
a rate $m \delta_{\text{D}}=\gamma$ and the total effective master
equation for the target systems of the distillation protocol is
given by
\begin{eqnarray*}
\dot{\rho}_{\mathcal{T}_{D}}=m \delta_{D}
E_{f_{\mathcal{T}_{\text{D}}}} (\rho )+\frac{\eps}{2} N(\rho).
\end{eqnarray*}
Hence, the resulting steady state fidelity is
\begin{eqnarray*}
f_{\mathcal{T}_{\text{D}}}=\left(f_{D}(f_{\mathcal{T}_{\text{sw}}})
\right)=\left(f_{D}(f_{\text{sw}}(f_I,\eps) ) \right),
\end{eqnarray*}
where $f_{D}(f_{\mathcal{T}_{\text{sw}}})$ is the entanglement
distilled from source systems with steady state fidelity
$f_{\mathcal{T}_{\text{sw}}}$.
In order to iterate this process, we require
$f_{\mathcal{T}_{\text{D}}}\geq f_I$, which can always be achieved
using a strong entanglement distillation (large $n$) and high
entangling rates $\gamma$ \cite{guu}.
\\The next iteration step begins with another continuous
entangling swapping procedure. Here, the target systems of the
distillation scheme act as source systems for the entanglement
swapping operation.
\\Since the total distance $L$ has been divided into $2^k$ segments (
$L=L_{0}2^k$), the protocol has to be iterated $k$ times. As
explained above, each iteration stage requires $2m^2n$ qubit pairs
such that in total $\left(2m^2n\right)^k$ source systems are
needed. Hence the resources scale with $(L/L_0)^{\log_2(2m^2n)}$
in the distance. We restrict the estimate of the required
resources to the number of used qubit pairs, since the other
resources scale polynomial in this quantity.
As specific example we consider the distribution of an entangled
state such that each repeater stage starts with and results in
links with fidelity $f=0.96$. We consider noise acing at a rate
$\epsilon=0.05$, distillation based on $n=16$ source systems (the
distillation protocol is described below) and stabilization of the
target pairs by means of $m=50$ copies of the underlying source
blocks and $\gamma\approx70$. In this example, the required
resources scale with
$(L/L_0)^{16.4}$.\\
\\
The entanglement distillation scheme used here is a four-to-one
distillation protocol for Werner states \cite{protocols} which is
applied two times in a nested fashion. Starting from four source
states $s_1$, $s_2$, $s_3$, $s_4$ with fidelity $f_{\text{in}}$,
the following operations are performed. First, a bilateral
$\text{CNOT}_{s_1\rightarrow s_2}$ operation is applied to the
first two pairs (where $s_1$ is the control and $s_2$ the target
qubit ) and $s_2$ is measured in the computational basis. Then, a
Hadamard transformation is performed on both qubits of $s_1$.
Subsequently, a bilateral $\text{CNOT}_{s_1\rightarrow s_3}$
operation is applied to the first and third pair (where $s_1$ is
the control and $s_3$ the target qubit) and $s_3$ is measured in
the computational basis. The measurement obtained on Alice's and
Bob's side are compared. If their measurement results coincide,
the resulting state $s_1$ is the desired higher entangled state.
If not, the "safety-copy" $s_4$ is used instead. In this event,
distillation was not successful and the fidelity has not been
increased, but in any case an entangled pair is available.
The fidelity of the resulting state is given by
\begin{eqnarray*}
f_{out}\left(f_{in}\right)=\frac{\left(1+g\right)\left(1+7g^2\right)}{16P_{\text{succ}}},
\end{eqnarray*}
where $g=\left(4f_{in}-1\right)/3$ and
$P_{\text{succ}}=\left(1+g^2+2g^3\right)/4$ is the success
probability of this protocol. In the example above, this protocol
is applied twice in a concatenated fashion. The second application
of the scheme is run using the output states of the first one as
input such that
$f_{D}(f_{in})=f_{out}\left(f_{out}\left(f_{in}\right)\right)$.\\
\\We conclude this section with an estimate of the convergence
speed of the presented repeater scheme. We start by considering
the elementary pairs constituting the entangled links on the
lowest level of the scheme. These systems reach the steady state
up to a certain high accuracy after a time $t_0$. After this time,
the dynamics of all systems on the next level is governed to a
good approximation by a time independent master equation and
converge with high accuracy to the steady state after another time
period of length $t_0$ has elapsed. Convergence of all $k$ levels
of the repeater scheme requires therefore a waiting time $k t_0$.
Since the number of levels used in the scheme scale only
logarithmical with the distance, we obtain a very moderate scaling
of the convergence time with the distance.
Note, that once the whole system is in a steady state, removal of
the final long-range entangled pair does not have an effect on the
underlying systems which remain in steady state.
The repeater protocol put forward here is based on continuous LOCC
maps, which represent a particular subset of possible dissipative
schemes. We have also presented a distillation protocol (scheme
I), without communication which does not fall in this class.
However, also the set of dissipative processes assisted by
classical communication includes other types of schemes not
covered here, which are yet to be explored.


\begin{thebibliography}{54}
\bibitem{HLOCC} LOCC channels are completely positive trace preserving maps that can be realized by means of Local Operations and Classical
 Communication.
\bibitem{HRetardationFootnote} We assume, that the time scales for classical communication $\Gamma^{-1}$ (see appendix) are sufficiently long
 such that retardation effects can be ignored.
\end{thebibliography}
\begin{thebibliography}{54}
\bibitem[S1]{FootnoteSchemeI}
 The variant of scheme I without communication is fundamentally
 different from scheme II, presented in Sec.~4 and cannot not be
 formulated in terms master equations of the form
 $\dot{\rho}\propto\left(T(\rho)-\rho\right)$, where $T(\rho)$ is a
 LOCC channel.
\bibitem[S3]{FootnoteSteadyStateError}
 The slowest term converges as $\mathcal{O}\left( \frac{t^3 \delta
 \Gamma^3}{(\Gamma+\delta)} e^{-t(\Gamma+\delta)} \right)$. After
 an initial waiting time of the order of $\frac{1}{\delta}$ the
 steady state is reached up to an error of the order
 $\mathcal{O}\left( \alpha ^{-2} e^{- \alpha^{-1}} \right)$.
\bibitem[S4]{Footnote3}
 It can be shown that this back-action has only a small effect on
 the steady state of the source system, and not accumulate if many
 source systems are coupled to $\T$. If the parameter $\delta$ is
 small, the back-action can be bounded several source system can be
 combined as described above. The strict avoidance of a state
 depended back-action is enforced only for the sake of clarity.
\bibitem[S7]{FootnoteSwapping}
 The channel $T_{\text{sw}}(\rho)$ corresponds to a three-partite
 LOCC map [11]. The results in Sec.~2 hold for bi-partite LOCC
 terms of the form $T(\rho)-\rho$ and can easily extended to
 multi-partite operations. The continuous realization of the swap
 scheme is already covered by the derivation for bi-partite LOCC
 operations presented above, since this protocol can be considered
 as an effective bi-partite scheme in which Alice and Charlie are
 treated as one party. Here, a classical one-way channel is used to
 send two copies of a message at the same time. One copy is sent to
 Alice and the other one is sent to Charlie.
\bibitem[S9]{guu}
 The error rate $\eps$ does not play an important role in this
 consideration, since the effect of noise on the target systems can
 be made negligible for large rates $\gamma$, without changing the
 scaling of the protocol. However, this comes at the cost of more
 resources, that are needed to achieve this rate $\gamma$ for the
 initial stage of the repeater protocol, i.e., the number of
 resources may be multiplied by a large factor, but the scaling
 with the length stays unchanged.
\end{thebibliography}
 \end{document}